\documentclass[aps,pre,twocolumn,groupedaddress]{revtex4}

\usepackage{graphicx,color}
\usepackage{bm}% bold math
\usepackage{amsmath}
\usepackage{verbatim}

\begin{document}
\title{Spectral solutions to stochastic models of gene expression with bursts and regulation}
\author{Andrew Mugler}
\email[]{ajm2121@columbia.edu}
\affiliation{Department of Physics, Columbia University, New York, NY 10027}
\author{Aleksandra M. Walczak}
\email[]{awalczak@princeton.edu}
\affiliation{Princeton Center for Theoretical Science, Princeton University, Princeton, NJ 08544}
\author{Chris H. Wiggins}
\email[]{chris.wiggins@columbia.edu}
\affiliation{Department of Applied Physics and Applied Mathematics, Center for Computational Biology and Bioinformatics, Columbia University, New York, NY 10027}

\date{\today}
\linespread{1}

\begin{abstract}
Signal-processing molecules inside cells are often present at low copy number, which necessitates probabilistic models to account for intrinsic noise.  Probability distributions have traditionally been found using simulation-based approaches which then require estimating the distributions from many samples.  Here we present in detail an alternative method for directly calculating a probability distribution by expanding in the natural eigenfunctions of the governing equation, which is linear.  We apply the resulting {\it spectral} method to three general models of stochastic gene expression: a single gene with multiple expression states (often used as a model of bursting in the limit of two states), a gene regulatory cascade, and a combined model of bursting and regulation.  In all cases we find either analytic results or numerical prescriptions that greatly outperform simulations in efficiency and accuracy.  In the last case, we show that bimodal response in the limit of slow switching is not only possible but optimal in terms of information transmission.

\end{abstract}

\maketitle

\newcommand{\bra}[1]{\langle{#1}|}
\newcommand{\ket}[1]{|{#1}\rangle{}}
\newcommand{\avg}[1]{\langle{#1}\rangle}
\newcommand{\bracket}[2]{\langle{#1}|{#2}\rangle{}}

\newcommand{\beqn}{\begin{eqnarray}}
\newcommand{\eeqn}{\end{eqnarray}}

\newcommand{\beq}{\begin{equation}}
\newcommand{\eeq}{\end{equation}}
\newcommand{\abs}[1]{|#1|}
\newcommand{\A}{{a^{\dagger}}}
\newcommand{\av}[1]{\langle{#1}\rangle{}}
\newcommand{\mk}[2]{\newcommand{#1}{#2}}
\newcommand{\rmk}[2]{\renewcommand{#1}{#2}}
\newcommand{\ip}[2]{\langle#1|#2\rangle}

\mk{\bbn}{b^+_nb^-_n}
\mk{\bbm}{b^+_mb^-_m}
\mk{\bbnn}{b^+_nb^-_n(n)}
\mk{\bbmn}{b_m^+b_m^-(n)}
\mk{\bb}{\bd_n\bl_n}
\mk{\bbbn}{\bd_n{\bar{b}^-_n}}
\mk{\bbbm}{\bd_m{\bar{b}^-_m}}
\mk{\bmb}{{\bar{\bl_m}}}
\mk{\bG}{{\bf{{G}}}}
\mk{\bmn}{\hat{b}_m^-(n)}
\mk{\bnn}{\hat{b}_n^-(n)}
\mk{\bbarn}{\hat{\bar{b}}_n^-}
\mk{\bbarm}{\hat{\bar{b}}_m^-}
\mk{\betan}{\hat{\beta}^-_n}
\mk{\betam}{\hat{\beta}^-_m}
\mk{\bbar}{\bar{b}^-}
\mk{\bd}{\hat{b}^+}
\mk{\ad}{\hat{a}^+}
\mk{\bl}{\hat{b}^-}
\mk{\al}{\hat{a}^-}
\mk{\qn}{q(n)}
\mk{\gn}{g(n)}
\mk{\qh}{\hat{q}_n}
\mk{\gh}{\hat{g}_n}
\mk{\qb}{\bar{q}}
\mk{\gb}{\bar{g}}
\mk{\Dn}{\hat{\Delta}_n}
\mk{\Gah}{\hat{\Gamma}_n}
\mk{\Lah}{\hat{\Lambda}_n}
\mk{\Deh}{\hat{\Delta}_n}
\mk{\Gjk}{G^{jk}}
\mk{\bj}{\<j|}
\mk{\bk}{\<k|}
\mk{\Gl}{{G_\ell}}
\mk{\Del}{\Delta}
\newcommand{\e}[1]{{\rm{e}}^{#1}}
\mk{\ehat}{\bf{\hat{ e}}}
\rmk{\L}{{\cal L}}
\rmk{\H}{\hat{H}}
\mk{\Om}{\hat{\Omega}}
\mk{\n}{\langle n\rangle}

Signals are processed in cells using networks of interacting components, including proteins, mRNAs, and small signaling molecules.  These components are usually present in low numbers \cite{Hooshangi, Thattai, Elowitz, Ozbudak, Swain, Paulsson2},
which means the size of the fluctuations in their copy counts is comparable to the copy counts themselves. Noise in gene networks has been shown to propagate \cite{Pedraza}, and therefore explicitly accounting for the stochastic nature of gene expression appears important when predicting the properties of real biological networks. 

Although summary statistics such as mean and variance are sometimes sufficient for answering questions of biological interest \cite{Paulsson}, calculating certain quantities, such as information transmission \cite{Tkacik, Tkacik2, Ziv, Mugler, Emberly, Tostevin} or escape properties \cite{Lan, Dellago, Dellago2, vanErp, vanErp2}, requires knowing the full probability distribution.
Full knowledge of the probability distribution can also be used to discern different molecular models of the noise sources based on recent exact measurements of probability distributions \cite{Elf, Choi, Raj, Golding}.

Much analytical and purely computational effort has gone into detailed models of noise in small genetic switches \cite{Paulsson, TanaseNicola, vanZon, VanZon2}. The most general description is based on the master equation describing the time evolution of the joint probability distribution over all copy counts \cite{vanKampen}. Some progress has been made by applying approximations to the master equation.  For example, a wide class of approximations focuses on limits of large concentrations or small switches \cite{Paulsson, TanaseNicola, Walczak2}. Approximations based on timescale separation of the steps of small signaling cascades have also been successfully used to calculate escape properties \cite{Lan, Lan2, Lan3}.  More often, modelers resort to stochastic simulation techniques, the most common of which is based on the varying-step Monte Carlo method \cite{Bortz, Gillespie}.  Probabilistic modeling of stochastic systems by simulation requires a computational challenge (generating many sample trajectories) followed by an even more difficult statistical challenge (parameterizing or otherwise estimating the probability distribution from which the samples are drawn) \cite{Bellman}.

In a recent paper \cite{Walczak} we introduce a new method for calculating the steady state distributions of chemical 
reactants, which we call the spectral method. The procedure relies on exploiting the natural basis of a simpler problem from the same class. The full problem is then solved numerically as an expansion in the natural basis
\footnote{A master equation in which the coordinate appears explicitly in the rates (e.g.\ $g_n$ or $q_n$ in Eqn.\ \ref{mastern}) is sometimes
mistakenly termed ``nonlinear'' in the literature, perhaps discouraging calculations which exploit
its inherent linear algebraic structure. We remind the reader that the master equation is perfectly linear.}.
In the spectral method we use the analytical guidance of a simple birth-death problem to reduce the master equation for a cascade to a set of algebraic equations. We break the problem into two parts: a parameter-independent preprocessing step, and the parameter-dependent step of obtaining the actual probability distributions. The spectral method allows huge computational gains with respect to simulations. In prior work \cite{Walczak} we illustrate the method in the example of gene regulatory cascades. We combine the spectral method with a Markov approximation, which exploits the observation that the behavior of a given species should depend only weakly on distant nodes given the proximal nodes.

In this paper we expand upon the application of the spectral method to more biologically realistic models of regulation: (i) a model of bursting in gene expression and (ii) a model that includes both bursts and explicit regulation by binding of transcription factor proteins. In both cases we demonstrate how the spectral method gives either analytic results or reduced algebraic expressions that can be solved numerically in orders of magnitude less time than stochastic simulations.  

We begin with a model of a multi-state birth-death process, a special case of which has been used to describe transcriptional bursting \cite{Raj, IyerBiswas}.  We illustrate how the spectral method reduces the model to a simple iterative algebraic equation, and in the appropriate limiting case recovers the known analytic results.  We also use this section to introduce the basic notation used throughout the paper.  Next we explore the problem of gene regulation in detail. The main idea behind the spectral method is the exploitation of an underlying natural basis for a problem which we can solve exactly.  We explore four different spectral representations of the regulation model used in previous work \cite{Walczak} that arise from four natural choices of eigenbasis in which to expand the solution (cf.\ Sec.\ \ref{bases} and Fig.\ \ref{diagram}).  All representations reduce the master equation to a set of linear algebraic equations, and one admits an analytic solution by virtue of the tridiagonal matrix algorithm.  We compare the efficiencies of the representations' numerical implementations and show that all outperform simulation.  Lastly, we apply the spectral method to a model that combines bursting and regulation.  We obtain a linear algebraic expression that permits large speedup over simulation and thus admits optimization of information transmission.  Optimization reveals two types of solutions: a unimodal response when the rates of switching between expression states are comparable to degradation rates, and a bimodal response when switching rates are much slower than degradation rates.

\section{Bursts of gene expression}
\label{burst}

We first consider a model of gene expression in which a gene exists in one of $Z$ stochastic ``states,''  i.e.\ protein production obeys a simple birth-death process, but with a state-dependent birth rate.  In the special case of $Z=2$, this corresponds to a gene existing in an on- or an off-state due, for example, to the binding and unbinding of the RNA polymerase.  Such a model has been used to describe transcriptional bursting \cite{Raj, IyerBiswas}, and we specialize to this case in Sec.\ \ref{onoff}.

For a general $Z$-state process, the master equation
\beq
\label{burstME}
\dot{p}_n^z = g_zp_{n-1}^z+(n+1)p_{n+1}^z-(g_z+n)p_n^z +\sum_{z'}\Omega_{zz'} p_n^{z'}
\eeq
describes the time evolution of the joint probability distribution $p_n^z$, where $z$ specifies the state ($1 \le z \le Z$), $n$ is the number of proteins, $g_z$ is the production rate in state $z$, $\Omega_{zz'}$ is a stochastic matrix of transition rates between states, and $\dot{p}_n^z$ denotes differentiation of the probability distribution with respect to time. Time and all rates have been nondimensionalized by the protein degradation rate.  Note that conservation of probability requires
\beq
\sum_z\Omega_{zz'}=0.
\eeq

The relationship between the transition rates in $\Omega_{zz'}$ and the probabilities $\pi_z = \sum_np_n^z$ of being in the $z$th state can be seen by summing Eqn.\ \ref{burstME} over $n$; at steady state one obtains
\beq
\label{pi1}
\sum_{z'}\Omega_{zz'}\pi_{z'}=0,
\eeq
and normalization requires
\beq
\label{pi2}
\sum_z\pi_z = 1.
\eeq
In the following sections, we introduce the spectral method and demonstrate how it can be used to solve for the full joint distribution $p_n^z$.

\subsection{Notation and definitions}
We begin our solution of Eqn.\ \ref{burstME} by defining the generating function \cite{vanKampen} $G_z(x)=\sum_np_n^zx^n$ over complex variable $x$
\footnote{Note that setting $x=e^{ik}$ makes clear that the generating function is simply the Fourier transform.}
(note that superscript $z$ is an index, while superscript $n$ on $x^n$ is a power).  It will prove more convenient to rewrite the generating function in a more abstract representation using states indexed by protein number $\ket{n}$,
\beq
\label{Gpz}
\ket{G_z}=\sum_np_n^z\ket{n},
\eeq
with inverse transform
\beq
\label{pGz}
p_n^z=\bracket{n}{G_z}.
\eeq
The more familiar form can be recovered by projecting the position space $\bra{x}$ onto Eqn.\ \ref{Gpz}, with the provision that
\beq
\label{xn}
\bracket{x}{n}=x^n.
\eeq
Concurrent choices of conjugate state
\beq
\label{nx}
\ip{n}{x}=\frac{1}{x^{n+1}}
\eeq
and inner product
\beq
\label{ip}
\ip{f}{f'}=\oint\frac{dx}{2\pi i}\ip{f}{x}\ip{x}{f'}
\eeq
ensure orthonormality of the states, $\ip{n}{n'}=\delta_{nn'}$, as can be verified using Cauchy's theorem,
\beq
\label{cauchy}
\oint \frac{dx}{2\pi i} \frac{f(x)}{(x-x_0)^{n+1}}
	= \frac{1}{n!} \partial^n_x \left[ f(x) \right]_{x=x_0} \theta(n+1),
\eeq
where the convention $\theta(0) = 0$ is used for the Heaviside function.

With these definitions, summing Eqn.\ \ref{burstME} over $n$ against $\ket{n}$ yields
\beq
\label{Gzop}
\ket{\dot{G}_z} = -(\ad-1)(\al-g_z)\ket{G_z}+\sum_{z'}\Omega_{zz'}\ket{G_{z'}},
\eeq
where the operators $\ad$ and $\al$ raise and lower protein number, respectively, i.e.\
\beqn
\label{ap}
\ad \ket{n}&=&\ket{n+1},\\
\label{am}
\al \ket{n}&=&n\ket{n-1},
\eeqn
with adjoint operations
\beqn
\label{apa}
\bra{n} \ad&=&\bra{n-1},\\
\label{ama}
\bra{n} \al&=&(n+1)\bra{n+1}.
\eeqn
As in the operator treatment of the simple harmonic oscillator in quantum mechanics, the raising and lower operators satisfy the commutation relation $[\al,\ad]=1$, and $\ad\al$ is a number operator, i.e.\ $\ad\al\ket{n} = n\ket{n}$
\footnote{Note here, however, the difference with respect to the normalization convention commonly used in quantum mechanics in the prefactors of the creation and annihilation operations.}.
This operator formalism for the generating function was introduced in the context of diffusion independently by Doi \cite{Doi} and Zeldovich \cite{Zeldovich}, and developed by Peliti \cite{Peliti}. A review by  Mattis and Glasser \cite{Mattis} introduces and discusses the applications of the formalism for diffusion.

The factorized form of the birth-death operator in Eqn.\ \ref{Gzop} suggests the definition of shifted raising and lowering operators
\beqn
\label{bpdef}
\bd &=& \ad-1\\
\label{bmdef}
\bl_z &=& \al-g_z,
\eeqn
making Eqn.\ \ref{Gzop}
\beq
\label{Gzop2}
\ket{\dot{G}_z} = -\bd\bl_z\ket{G_z}+\sum_{z'}\Omega_{zz'}\ket{G_{z'}}.
\eeq
Since $\bd\bl_z$ is a new number operator, it is clear that the eigenvalues of the birth-death operator $-\bd\bl_z$ are nonpositive integers, i.e.\
\beq
\bd\bl_z\ket{j_z} = j\ket{j_z},
\eeq
where nonnegative integers $j$ index ($z$-dependent) eigenfunctions $\ket{j_z}$.  In position space the $j$th eigenfunction is
\beq
\label{jket}
\ip{x}{j_z} = (x-1)^j e^{g_z(x-1)},
\eeq
with conjugate
\beq
\label{jbra}
\ip{j_z}{x} = \frac{e^{-g_z(x-1)}}{(x-1)^{j+1}},
\eeq
such that orthonormality $\ip{j_z}{j'_z} = \delta_{jj'}$ is satisfied under the inner product in Eqn.\ \ref{ip}.  Note that $\bd$ and $\bl_z$ raise and lower eigenstates $\ket{j_z}$ as in Eqns.\ \ref{ap}-\ref{ama}, i.e.\
\beqn
\label{bp}
\bd \ket{j_z}&=&\ket{(j+1)_z},\\
\label{bm}
\bl_z \ket{j_z}&=&j\ket{(j-1)_z},\\
\label{bpa}
\bra{j_z} \bd&=&\bra{(j-1)_z},\\
\label{bma}
\bra{j_z} \bl_z&=&(j+1)\bra{(j+1)_z}.
\eeqn

As we will see in this and subsequent sections, going between the protein number basis $\ket{n}$ and the eigenbasis $\ket{j_z}$ requires the mixed product $\ip{n}{j_z}$ or its conjugate $\ip{j_z}{n}$.  There are several ways of computing these objects, as described in Appendix A.  Notable special cases are
\beqn
\label{0_on_n}
\ip{(j=0)_z}{n} &=& 1,\\
\label{n_on_0}
\ip{n}{(j=0)_z} &=& e^{-g_z}\frac{(g_z)^n}{n!},
\eeqn
where the latter is the Poisson distribution.

\subsection{The spectral method}
\label{specz}

We now demonstrate how the spectral method exploits the eigenfunctions $\ket{j_z}$ to decompose and simplify the equation of motion.  Expanding the generating function in the eigenbasis,
\beq
\label{expandz}
\ket{G_z} = \sum_jG_j^z\ket{j_z},
\eeq
and projecting the conjugate state $\bra{j_z}$ onto Eqn.\ \ref{Gzop2} yields the equation of motion
\beq
\label{cEoM}
\dot{G}_j^z = -jG_j^z + \sum_{z'}\Omega_{zz'}\sum_{j'}G_{j'}^{z'} \ip{j_z}{j'_{z'}}
\eeq
for the expansion coefficients $G_j^z$ (where the dummy index $j$ in Eqn.\ \ref{expandz} has been changed to $j'$ in Eqn.\ \ref{cEoM}).  Using Eqns.\ \ref{ip}, \ref{cauchy}, \ref{jket}, and \ref{jbra}, the product $\ip{j_z}{j'_{z'}}$ evaluates to
\beq
\label{overlap}
\ip{j_z}{j'_{z'}} = \frac{(-\Delta_{zz'})^{j-j'}}{(j-j')!}\theta(j-j'+1),
\eeq
where $\Delta_{zz'} = g_z - g_{z'}$.  Noting that $\ip{j_z}{j_{z'}}=1$ and that $\ip{j_z}{j'_{z}}=0$ for $j'<j$, Eqn.\ \ref{cEoM} becomes
\beqn
\dot{G}_j^z &=& -jG_j^z + \sum_{z'}\Omega_{zz'}G_j^{z'}\nonumber\\
\label{cEoM2}
&&	+\sum_{z'\ne z}\Omega_{zz'}\sum_{j'<j}G_{j'}^{z'}\frac{(-\Delta_{zz'})^{j-j'}}{(j-j')!}.
\eeqn
The last term in Eqn.\ \ref{cEoM2} makes clear that each $j$th term is slaved to terms with $j'<j$, allowing the $G_j^z$ to be computed iteratively in $j$.  The lower-triangular structure of the equation is a consequence of rotating to the eigenspace of the birth-death operator; this structure was not present in the original master equation.

At steady-state, $G_j^z$ obeys
\beq
\label{css}
jG_j^z - \sum_{z'}\Omega_{zz'}G_j^{z'} = \sum_{z'\ne z}\Omega_{zz'}\sum_{j'<j}G_{j'}^{z'}\frac{(-\Delta_{zz'})^{j-j'}}{(j-j')!},
\eeq
from which it is clear that $G_j^z$ can be computed successively in $j$.  Since
\beq
G_0^z = \ip{(j=0)_z}{G_z} = \sum_np_n^z\ip{(j=0)_z}{n} = \pi_z
\eeq
(cf.\ Eqn.\ \ref{0_on_n}), the computation is initialized using Eqns.\ \ref{pi1}-\ref{pi2}, i.e.\
\beqn
\label{init1}
\sum_{z'}\Omega_{zz'}G_0^{z'} &=& 0,\\
\label{init2}
\sum_z G_0^z &=& 1.
\eeqn
Recalling Eqns.\ \ref{pGz} and \ref{expandz}, the probability distribution is retrieved via
\beq
p_n^z = \sum_jG_j^z\ip{n}{j_z},
\eeq
where the mixed product $\ip{n}{j_z}$ can be computed as described in Appendix A.

There is an alternative way to decompose the master equation spectrally.  Instead of expanding the generating function in eigenfunctions $\ket{j_z}$, which depend on the production rates $g_z$ in each state, we may expand in eigenfunctions parameterized by a single rate $\gb$, i.e.\
\beq
\label{expandbar}
\ket{G_z} = \sum_jG_j^z\ket{j},
\eeq
where
\beqn
\label{jbarket}
\ip{x}{j} &=& (x-1)^j e^{\gb(x-1)},\\
\label{jbarbra}
\ip{j}{x} &=& \frac{e^{-\gb(x-1)}}{(x-1)^{j+1}}.
\eeqn
The parameter $\gb$ is arbitrary and thus acts as a ``gauge'' freedom.

We may now partition the birth-death operator as
\beq
-\bd\bl_z = -\bd\bbar-\bd\Gamma_z
\eeq
where $\bbar=\al-\gb$ such that the $\ket{j}$ are the eigenstates of $\bd\bbar$, i.e.\
\beq
\bd\bbar\ket{j} = j\ket{j},
\eeq
and $\Gamma_z = \gb-g_z$ describes the deviation of each state's production rate from the constant $\gb$.

Projecting the conjugate state $\bra{j}$ onto Eqn.\ \ref{Gzop2} and using Eqn.\ \ref{expandbar} (with dummy index $j$ changed to $j'$) gives
\beq
\label{cEoMbar}
\dot{G}_j^z = -jG_j^z + \sum_{j'}\bra{j}\bd\Gamma_z\ket{j'}G_{j'}^z+\sum_{z'}\Omega_{zz'}G_{j}^{z'}.
\eeq
Recalling Eqn.\ \ref{bpa}, Eqn.\ \ref{cEoMbar} at steady state becomes
\beq
\label{cEoMbar2}
\sum_{z'}\left(\Omega_{zz'}-j\delta_{zz'}\right)G_{j}^{z'} = \Gamma_zG_{j-1}^z.
\eeq
Eqn.\ \ref{cEoMbar2} is subdiagonal in $j$, meaning computation of the $j$th term requires only the previous $(j-1)$th term and the inversion of the $Z$-by-$Z$ matrix $({\bf \Omega}-j{\bf I})$ (where ${\bf I}$ is the identity matrix).  It is initialized with Eqns.\ \ref{init1}-\ref{init2} and solved successively in $j$.  The probability distribution is retrieved via
\beq
p_n^z = \sum_jG_j^z\ip{n}{j},
\eeq
where $\ip{n}{j}$ is computed as described in Appendix A.

\begin{figure}
\begin{center}
\includegraphics[scale=0.47]{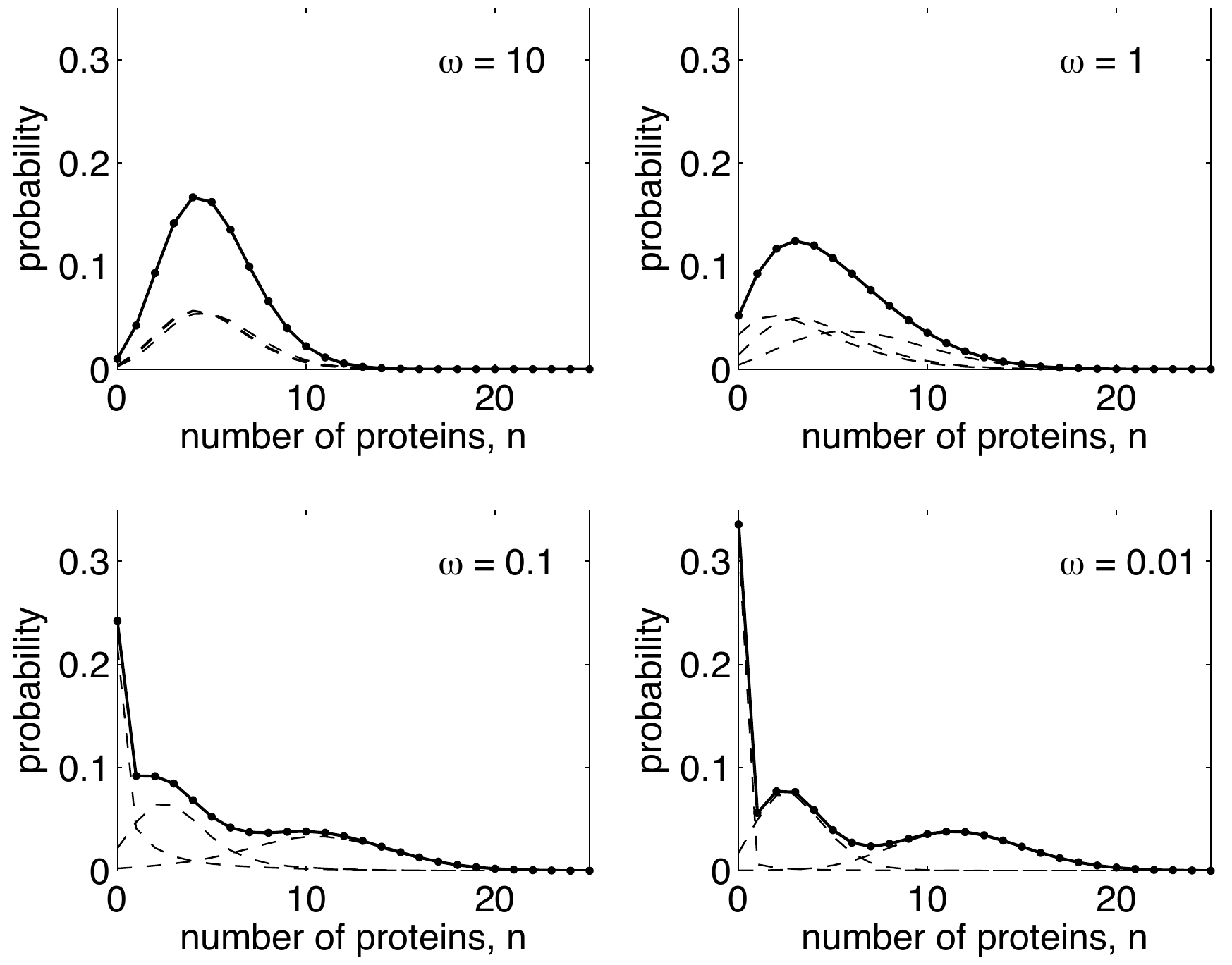}
\linespread{1}
\caption{An example of the spectral method for $Z=3$ states.  Dotted curves show the joint distribution $p_n^z$ for each of $z=1,2,$ and $3$, and solid curves show the marginal distribution $p_n$.  The stochastic transition matrix is given in Eqn.\ \ref{Omega3}, and the setting of $\omega$ in each panel is indicated in the upper-right corner.  Production rates are $g_1=0$, $g_2=3$, and $g_3=12$ for all panels.  Distributions are calculated via the spectral decomposition in Eqn.\ \ref{cEoMbar2} with $\gb=\langle g_z\rangle = 5$.}
\label{3state}
\end{center}
\end{figure}

As an example of a simple computation employing the spectral method, Fig.\ \ref{3state} shows probability distributions for the case of $Z=3$ states, corresponding to a gene that is either off, producing proteins at a low rate, or producing proteins at a high rate.  For simplicity we set the rates of switching among all states equal to a constant $\omega$, making the stochastic matrix
\beq
\label{Omega3}
{\bf \Omega} = \begin{pmatrix} -2\omega & \omega & \omega \\
	\omega & -2\omega & \omega \\
	\omega & \omega & -2\omega \end{pmatrix}.
\eeq
As seen in Fig.\ \ref{3state}, when $\omega\ll 1$ (corresponding to a switching rate much slower than the degradation rate) the dwell time in each expression state lengthens.  The slow switching gives the protein copy number time to equilibrate in any of the three expression states, resulting in a trimodal marginal distribution $p_n$.   When $\omega\gg 1$ (corresponding to a switching rate much faster than the degradation rate), the system switches frequently among the three expression states, resulting in an average production rate.  In this limit, the expression state equilibrates on a faster timescale than the protein number state.

\subsection{The on/off gene}
\label{onoff}

For the special case of $Z=2$ states, as when a gene is either ``on'' ($z=+$) or ``off'' ($z=-$), it is useful to demonstrate how the spectral method reproduces known analytic results \cite{Raj, IyerBiswas}.  The probability distribution can be written in vector form as
\beq
\vec{p}_n = \begin{pmatrix} p_n^- \\ p_n^+ \end{pmatrix},
\eeq
and defining $\omega_+$ and $\omega_-$ as the transition rates to and from the on-state, respectively, the stochastic matrix takes the form
\beq
\label{Omega2}
{\bf \Omega} = \begin{pmatrix} -\omega_+ & \omega_- \\ \omega_+ & -\omega_- \end{pmatrix}.
\eeq
Note that Eqn.\ \ref{pi1} implies
\beq
\label{piom}
\frac{\pi_-}{\pi_+} = \frac{\omega_-}{\omega_+},
\eeq
which makes clear that increasing the rate of transition to either state increases the probability of being in that state.

From Eqn.\ \ref{css} the spectral expansion coefficients obey
\beq
jG_j^\pm+\omega_\mp G_j^\pm-\omega_\pm G_j^\mp = \omega_\pm\sum_{j'<j}G_{j'}^\mp\frac{(\mp\Delta)^{j-j'}}{(j-j')!},
\eeq
where $\Delta = \Delta_{+-} = -\Delta_{-+}$.  Initializing with $G_0^\pm = \omega_\pm/(\omega_++\omega_-)$ and computing the first few terms reveals the pattern
\beqn
G_j^\pm = \frac{\omega_\pm}{\omega_++\omega_-}\frac{(\mp\Delta)^j}{j!}\frac{\prod_{j'=0}^{j-1}(j'+\omega_\mp)}{\prod_{j''=0}^{j-1}(j''+\omega_++\omega_-+1)}\quad&&\\
= \frac{\omega_\pm}{\omega_++\omega_-}\frac{(\mp\Delta)^j}{j!}\frac{\Gamma(j+\omega_\mp)}{\Gamma(\omega_\mp)}\frac{\Gamma(\omega_++\omega_-+1)}{\Gamma(j+\omega_++\omega_-+1)},
\quad&&
\eeqn
where in the second line the products are written in terms of the Gamma function.

For comparison with known results \cite{Raj, IyerBiswas} we write the total generating function $\ket{G}=\sum_\pm\ket{G_\pm}$ in position space,
\beqn
G(x) &=& \sum_\pm\ip{x}{G_\pm} = \sum_\pm\sum_j\ip{x}{j_\pm}\ip{j_\pm}{G_\pm}\\
&=& \sum_\pm\sum_j(x-1)^j e^{g_\pm(x-1)}G_j^\pm\\
&=& \sum_\pm \frac{\omega_\pm}{\omega_++\omega_-}e^{g_\pm(x-1)}\nonumber\\
\label{Ghyp}
&&\times\Phi[\omega_\mp,\omega_++\omega_-+1;\mp\Delta(x-1)],
\eeqn
where
\beq
\label{hyp}
\Phi[\alpha,\beta;y] = \sum_j\frac{\Gamma(j+\alpha)}{\Gamma(\alpha)}\frac{\Gamma(\beta)}{\Gamma(j+\beta)}\frac{y^j}{j!}
\eeq
is the confluent hypergeometric function. As shown in Appendix B, in the limit $g_-=0$, Eqn.\ \ref{Ghyp} reduces to
\beq
\label{sri}
G(x) = \Phi[\omega_+,\omega_++\omega_-;g_+(x-1)],
\eeq
and the marginal $p_n$ is given by
\beqn
p_n &=& \frac{g_+^n}{n!}\frac{\Gamma(n+\omega_+)}{\Gamma(\omega_+)}\frac{\Gamma(\omega_++\omega_-)}{\Gamma(n+\omega_++\omega_-)}\nonumber\\
\label{raj}
&&\times\Phi[\omega_++n,\omega_++\omega_-+n;-g_+],
\eeqn
in agreement with the results of Iyer-Biswas et al.\ \cite{IyerBiswas} and Raj et al.\ \cite{Raj}.
We remind the reader that in addition to reducing to known results in the special case of $Z=2$ states with a vanishing off-rate, the spectral method is valid for any number of states with arbitrary production rates.

\section{Gene regulation}
\label{reg}

Next we consider a two-gene regulatory cascade, in which the production rate of the second gene is a function of the number of proteins of the first gene. As shown in previous work \cite{Walczak}, a cascade of any length can be reduced to such a generalized two-dimensional system using the Markov approximation, which asserts that the probability distribution for a given node of the cascade should depend only weakly on the probability distributions of the distant modes given the proximal nodes. 

In the present section, we consider only the generalized two-dimensional equation and explore different approaches to solving it.
The equation describes two genes, each with one expression state, with regulation encoded by a functional dependence of the downstream protein production rate on the upstream protein copy number. In section \ref{threshonoff} we make an explicit connection between the on/off gene discussed in section \ref{onoff} and the case when the functional dependence is a threshold. Finally, in section \ref{burstreg} we combine the two types of models and consider a system with regulation and bursts.

\subsection{Representations of the master equation}
\label{bases}
\subsubsection{The $\ket{n,m}$ basis}

\begin{figure}
\begin{center}
\includegraphics[scale=.7]{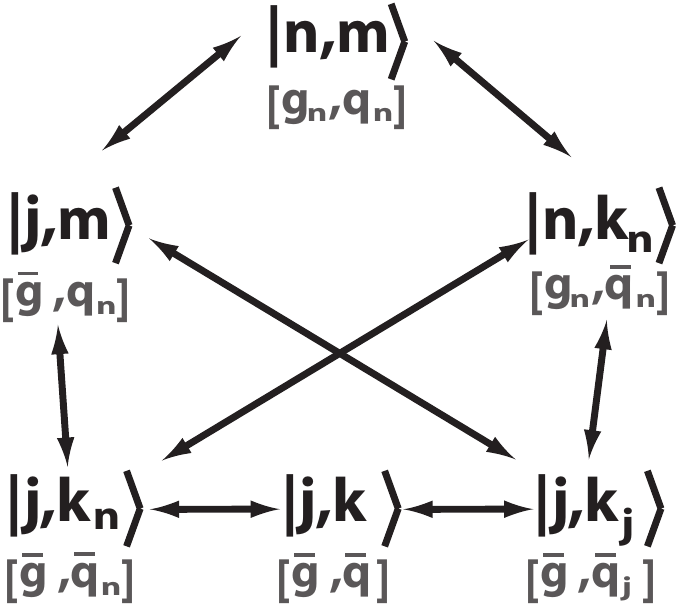}
\linespread{1}
\caption{Summary of the bases discussed in Sec.\ \ref{bases} (black) and their gauge freedoms (barred parameters in grey).  In the top row, neither the $m$ nor the $n$ sector is expanded in an eigenbasis; in the middle row, one sector is expanded; and in the bottom row, both sectors are expanded.  The
$\ket{j,k}$ basis can be viewed as a special case of the $\ket{j,k_j}$ or $\ket{j,k_n}$ basis with $\qb_j=\qb$ or $\qb_n=\qb$ respectively.  The $\ket{j,m}$ basis is not discussed as it is not useful in simplifying the problem.}
\label{diagram}
\end{center}
\end{figure}

Defining $n$ and $m$ as the numbers of proteins produced by the first and second gene, respectively, the master equation describing the time evolution of the joint probability distribution $p_{nm}$ is \cite{Walczak}
\beqn
\label{mastern}
\dot{p}_{nm}=g_{n-1}p_{n-1,m}+(n+1)p_{n+1,m}-(g_n+n)p_{nm}&&\nonumber\\
	+\rho\left[q_np_{n,m-1}+(m+1)p_{n,m+1}-(q_n+m)p_{nm}\right].\quad&&
\eeqn
The function $q_n$ describes the regulation of the second species by the first, and the function $g_n$ describes the effective autoregulation of the first species, due either to a non-Poissonian input distribution or to effects further upstream in the case of a longer cascade \cite{Walczak}.  Time is rescaled by the first gene's degradation rate, so that each gene's production rate ($g_n$ or $q_n$) is normalized by its respective degradation rate, and $\rho$ is ratio of the second gene's degradation rate to that of the first. We impose no constraints on the form of $g_n$ or $q_n$---they can be arbitrary nonlinear functions.

Summing Eqn.\ \ref{mastern} over $m$ gives a simple recursion relation between $g_n$ and $p_n$ at steady state, from which explicit relations are easily identified.  If $p_n$ is known, $g_n$ is found as
\beq
\label{gn}
g_n = (n+1)\frac{p_{n+1}}{p_n}.
\eeq
If on the other hand $g_n$ is known, $p_n$ is found as
\beq
\label{pn}
p_n = \frac{p_0}{n!}\prod_{n'=0}^{n-1}g_{n'},
\eeq
with $p_0$ set by normalization.  Note that if the first species obeys a simple birth-death process, $g_n=g=$ constant, and Eqn.\ \ref{pn} reduces to the Poisson distribution.

In the current representation (Eqn.\ \ref{mastern}), which we denote the $\ket{n,m}$ basis, finding the steady state solution for the joint distribution $p_{nm}$ means finding the null space of
an infinite (or, effectively for numerical purposes, very large) locally banded tridiagonal matrix.
More precisely, defining $N$ as the numerical cutoff in protein number $n$ or $m$, the problem amounts to inverting an $N^2$-by-$N^2$ matrix, which is computationally taxing even for moderate cutoffs $N$.

In order to solve Eqn \ref{mastern} more efficiently we will employ the spectral method.
We begin as before by defining the generating function \cite{vanKampen} $G(x,y)=\sum_{nm}p_{nm}x^ny^m$ over complex variables $x$ and $y$,
or, in state notation,
\beq
\label{Gp}
\ket{G}=\sum_{nm}p_{nm}\ket{n,m},
\eeq
with inverse transform
\beq
\label{pG}
p_{nm}=\bracket{n,m}{G}.
\eeq
Summing Eqn.\ \ref{mastern} over $n$ and $m$ against $\ket{n,m}$ and employing the same operator notation as in Eqns.\ \ref{ap}-\ref{ama} yields
\beq
\label{eom}
\dot{\ket{G}}=-\H\ket{G},
\eeq
where
\beq
\label{H}
\H=\bd_n\bnn+\rho\bd_m\bmn,
\eeq
and
\beqn
\label{b1}
\bd_n &=& \ad_n-1,\\
\label{b2}
\bd_m &=& \ad_m-1,\\
\label{b3}
\bnn &=& \al_n - \gh,\\
\label{b4}
\bmn &=& \al_m - \qh.
\eeqn
Here the regulation functions have been promoted to operators obeying $\gh\ket{n}=g_n\ket{n}$ and $\qh\ket{n}=q_n\ket{n}$, subscripts on operators denote the sector ($n$ or $m$) on which they operate, and the arguments of $\bl_n$ and $\bl_m$ remind us that both are $n$-dependent.

Eqn.\ \ref{H} makes clear that if not for the $n$-dependence of the operators the Hamiltonian $\H$ would be diagonalizable, or, equivalently, if $g_n$ and $q_n$ were constants the master equation would factorize into two individual birth-death processes. 
We may still, however, partition the full Hamiltonian as
\beqn
\label{bpeqn}
\H&=&\H_0+\H_1,
\eeqn
where $\H_0$ is a diagonalizable part (and $\H_1$ is the corresponding deviation from the diagonal form), and expand $\ket{G}$ in the eigenbasis of $\H_0$ to exploit the diagonality.

As with the multi-state system in Sec.\ \ref{specz}, where we expand the solution in two different bases, there are several natural choices of eigenbasis of $\H_0$.  Fig.\ \ref{diagram} summarizes these choices diagrammatically: starting in the $\ket{n,m}$ basis (at the top of Fig.\ \ref{diagram}), one may expand in eigenfunctions either the first species, yielding the $\ket{j,m}$ basis (left), or the second species, yielding the $\ket{n,k_n}$ basis (right; in general we allow the parameter defining the second species' eigenfunctions to depend on $n$ to reflect the regulation of the second species by the first
From either the $\ket{j,m}$ or the $\ket{n,k_n}$ basis, one may expand in eigenfunctions the remaining species, yielding either the $\ket{j,k_n}$ basis (bottom left), in which the second species' eigenfunctions depend on the first species' copy number $n$, or the $\ket{j,k_j}$ basis (bottom right), in which the second species' eigenfunctions depend on the first species' eigenmode number $j$.  Both bases reduce to the $\ket{j,k}$ basis (bottom center) when the parameter of the second species' eigenfunctions is a constant.

The $\ket{n,m}$ and $\ket{j,m}$ bases are less numerically useful than the other bases: as discussed above and detailed in Fig.\ \ref{fast}, the $\ket{n,m}$ basis is numerically inefficient; and the $\ket{j,m}$ basis does not exploit the natural structure of the problem, 
since, unlike the other bases, it neither retains the tridiagonal structure in $n$ nor gains a lower triangular structure in $k$ (see the sections below).  Each of the remaining bases, however, has preferable properties in terms of numerical stability and
ability to represent the function sparsely yet accurately (either using a few
values of n in the $\ket{n,k_n}$ basis or a few values of j in the $\ket{j,k}$, $\ket{j,k_j}$, or $\ket{j,k_n}$ bases, for example). Moreover, the equation of motion
simplifies differently in each of the different bases.  We present the derivations of the equations of motion in the following sections, beginning with the $\ket{j,k}$ basis, generalizing to the $\ket{j,k_j}$ basis, moving to the $\ket{n,k_n}$ basis, and ending with the $\ket{j,k_n}$ basis.

 \subsubsection{The $\ket{j,k}$ basis}
\label{jkbasis}

For expository purposes we start by recalling the spectral representation used in previous work \cite{Walczak}. We choose the diagonal part of the Hamiltonian to correspond to two birth-death process with constant production rates $\gb$ and $\qb$,
\beq
\label{H0_1}
\H_0=\bd_n\bbar_n+\rho\bd_m\bbar_m,
\eeq
where $\bbar_n = \al_n - \gb$ and $\bbar_m =  \al_m - \qb$.
As in Sec.\ \ref{specz}, the parameters $\gb$ and $\qb$ act as gauge choices: their values can be set arbitrarily and thus can affect the numerical stability of the method.
The nondiagonal part,
\beq
\label{H1_1}
\H_1= \bd_n\Gah + \rho\bd_m\Deh,
\eeq
captures the deviations $\Gah = \gb-\gh$ and $\Deh = \qb-\qh$
of the regulation functions from the constant rates.
We expand the generating function as
\beq
\label{G_1}
\ket{G} = \sum_{jk} G_{jk} \ket{j,k},
\eeq
where $\ket{j,k}$ is the eigenbasis of $\H_0$, i.e.\
\beq
\H_0 \ket{j,k}=(j+\rho k)\ket{j,k}.
\eeq
The eigenbasis is parameterized by the rates $\gb$ and $\qb$, meaning in position space $\ip{x}{j}$ is as in Eqn.\ \ref{jbarket} and similarly for $\ip{y}{k}$ with $x\rightarrow y$, $j\rightarrow k$, and $\gb\rightarrow \qb$.

With Eqns.\ \ref{H0_1}-\ref{G_1}, projecting the conjugate state $\bra{j,k}$ onto Eqn.\ \ref{eom} yields
\beqn
\dot{G}_{jk} &=& -(j+\rho k)G_{jk}-\sum_{j'k'}\bra{j}\bd_n\Gah\ket{j'}\ip{k}{k'}G_{j'k'}\nonumber\\
\label{jk_temp}
&&	-\rho\sum_{j'k'}\bra{k}\bd_m\ket{k'}\bra{j}\Deh\ket{j'}G_{j'k'}
\eeqn
(where the dummy indices $j$ and $k$ in Eqn.\ \ref{G_1} have been changed to $j'$ and $k'$ respectively in Eqn.\ \ref{jk_temp}).
Recalling Eqn.\ \ref{bpa} and restricting attention to steady state, Eqn.\ \ref{jk_temp} becomes
\beqn
0&=&-(j+\rho k)G_{jk}-\sum_{j'}\Gamma_{j-1,j'}G_{j'k}\nonumber\\
\label{jk_final}
&&	-\rho\sum_{j'}\Delta_{jj'}G_{j',k-1},
\eeqn
where the deviations have been rotated into the eigenbasis as
\beqn
\label{Gammajj}
\Gamma_{jj'}&=&\bra{j}\Gah\ket{j'}=\sum_{n}\ip{j}{n}(\gb-g_n)\ip{n}{j'},\\
\Delta_{jj'}&=&\bra{j}\Deh\ket{j'}=\sum_{n}\ip{j}{n}(\qb-q_n)\ip{n}{j'}.
\eeqn

Eqn.\ \ref{jk_final} is subdiagonal in $k$ and is therefore similar to Eqn.\ \ref{cEoMbar2} in that the last term acts as a source term.  It is initialized using
\beqn
G_{j0}&=&\ip{j,k=0}{G}=\sum_{nm}p_{nm}\ip{j}{n}\ip{0}{m}\\
\label{Ginit}
&=&\sum_np_n\ip{j}{n},
\eeqn
(cf.\ Eqn.\ \ref{0_on_n}) with known $p_n$ (cf.\ Eqn.\ \ref{pn}), then solved at each subsequent $k$ using the result for $k-1$.  Eqn.\ \ref{jk_final} can be written in linear algebraic notation as
\beq
\label{jk_mat}
\vec{G}_k = -\rho\left({\bf D}^k+{\bf S}^-{\bf \Gamma}\right)^{-1}{\bf \Delta}\vec{G}_{k-1},
\eeq
where $\vec{G}_k$ is a vector over $j$, bold denotes matrices, and $D_{jj'}^k=(j+\rho k)\delta_{jj'}$ and $S_{jj'}^-=\delta_{j-1,j'}$
are diagonal and subdiagonal matrices, respectively.  Eqn.\ \ref{jk_mat} makes clear that the solution involves only matrix multiplication and the inversion of a $J$-by-$J$ matrix $K$ times, where $J$ and $K$ are cutoffs in the eigenmode numbers $j$ and $k$, respectively.  In fact, if the first species obeys a simple birth-death process, i.e.\ $g_n=g=$ constant, setting $\gb=g$ makes $\Gamma_{jj'}=0$, and, since ${\bf D}^k$ is diagonal, the solution involves only matrix multiplication.  The decomposition of the master equation into a linear algebraic equation results in huge gains in efficiency over direct solution in the $\ket{n,m}$ basis; the efficiency of all bases presented in this section is described in Sec.\ \ref{compare} and illustrated in Fig.\ \ref{fast}.

Recalling Eqns.\ \ref{pG} and \ref{G_1}, the joint distribution is retrieved from $G_{jk}$ via the inverse transform
\beq
p_{nm}=\sum_{jk} \bracket{n}{j} G_{jk} \bracket{m}{k},
\eeq
a computation again involving only matrix multiplication.  The mixed product matrices $\ip{n}{j}$ and $\ip{m}{k}$ are computed as described in Appendix A.

\subsubsection{The $\ket{j,k_j}$ basis}
The $\ket{j,k}$ basis treats both genes similarly by expanding each around a constant production rate.  We may instead imagine an eigenbasis that more closely reflects the underlying asymmetry imposed by the regulation, and make the basis of the second gene a function of that of the first.  That is, we expand the first gene in a basis $\ket{j}$ with gauge $\gb$ as before, but now we expand the second gene in a basis $\ket{k_j}$ with a $j$-dependent local gauge $\qb_j$.
We write the generating function as
\beq
\label{G_2}
\ket{G} = \sum_{jk} G_{jk} \ket{j,k_j},
\eeq
and Eqn.\ \ref{eom} at steady state becomes
\beq
\label{eom_2}
0 = -\sum_{jk} G_{jk} \left[ \H_0(j)+\H_1(j)\right]\ket{j,k_{j}},
\eeq
where we have partitioned the Hamiltonian for each $j$ as
\beqn
\label{H0_2}
\H_0(j)&=&\bd_n\bbar_n+\rho\bd_m\bbar_m(j),\\
\label{H1_2}
\H_1(j)&=&\bd_n\Gah + \rho\bd_m\Deh(j),
\eeqn
with $\bbar_n = \al_n - \gb$ and $\Gah = \gb-\gh$ as before and now $\bbar_m(j) =  \al_m - \qb_j$ and $\Deh(j) = \qb_j-\qh$.  Note that this basis enjoys the eigenvalue equation
\beq
\H_0(j)\ket{j,k_j} = (j+\rho k)\ket{j,k_j}.
\eeq

Projecting the conjugate state $\bra{j,k_j}$ onto Eqn.\ \ref{eom_2} yields, after some simplification (cf.\ Appendix C), the equation of motion
\beqn
&0& =(j+\rho k)G_{jk}+\sum_{j'}\Gamma_{j-1,j'}G_{j'k}\nonumber\\
\label{jkj_final}
&&	+\sum_{\ell=1}^k\sum_{j'}\left( \Gamma_{j-1,j'}V^\ell_{jj'}
	+\rho\Delta_{jj'}V^{\ell-1}_{jj'}\right)G_{j',k-\ell},\qquad
\eeqn
where $\Gamma_{jj'}$ is as in Eqn.\ \ref{Gammajj} and
\beqn
\Delta_{jj'}&=&\bra{j}\Deh(j)\ket{j'}=\sum_{n}\ip{j}{n}(\qb_j-q_n)\ip{n}{j'},\qquad\\
V^\ell_{jj'} &=& \frac{(-Q_{jj'})^\ell}{\ell!},
\eeqn
with $Q_{jj'}=\qb_j-\qb_{j'}$.
Eqn.\ \ref{jkj_final} can be written linear algebraically as
\beqn
\vec{G}_k&=&-\left({\bf D}^k+{\bf S}^-{\bf \Gamma}\right)^{-1}\nonumber\\
\label{jkj_mat}
&&	\times\sum_{\ell=1}^k\left[({\bf S}^-{\bf \Gamma})*{\bf V}^\ell
	+\rho{\bf \Delta}*{\bf V}^{\ell-1}\right]\vec{G}_{k-\ell},\qquad
\eeqn
where $D^k_{jj'}$ and $S^-_{jj'}$ are defined as before (cf.\ Eqn.\ \ref{jk_mat}), and $*$ denotes an element-by-element matrix product.  Once again, Eqn.\ \ref{jkj_mat} is lower-triangular in $k$ and requires only matrix multiplication and the inversion of a $J$-by-$J$ matrix $K$ times.  $\vec{G}_k$ is initialized as in Eqn.\ \ref{Ginit}, and the $k$th term is computed from the previous $k'<k$ terms.
The joint distribution is retrieved via the inverse transform
\beq
p_{nm}=\sum_{jk} \bracket{n}{j} G^{jk} \bracket{m}{k_j}.
\eeq

\subsubsection{The $\ket{n,k_n}$ basis}

Expansion in either the $\ket{j,k}$ or the $\ket{j,k_j}$ basis conveniently turns the master equation into a lower-triangular linear algebraic equation and replaces cutoffs in protein number with cutoffs in eigenmode number (which can be smaller with appropriate choices of gauge).  However, these bases sacrifice the original tridiagonal structure of the master equation in the copy number of first gene, $n$.  Therefore we now consider a mixed representation, in which the first gene remains in protein number space $\ket{n}$, and we expand the second gene in an $n$-dependent eigenbasis $\ket{k_n}$,
\beq
\ket{G} = \sum_{nk} G_{nk} \ket{n,k_n}.
\eeq
If the rate parameter of the $\ket{k_n}$ basis were the regulation function $q_n$, it would be natural to make $\H_0$ the entire Hamiltonian, Eqn.\ \ref{H}.  For generality we will instead allow the rate parameter of the $\ket{k_n}$ basis to be an arbitrary $n$-dependent local gauge $\qb_n$, such that Eqn.\ \ref{eom} at steady state naturally partitions as
\beq
\label{eom_3}
0 = -\sum_{nk} G_{nk} \left[ \H_0(n)+\H_1(n)\right]\ket{n,k_n},
\eeq
where
\beqn
\label{H0_3}
\H_0(n)&=&\bd_n\bl_n+\rho\bd_m\bbar_m(n),\\
\label{H1_3}
\H_1(n)&=&\rho\bd_m\Deh(n),
\eeqn
with $\bbar_m(n) =  \al_m - \qb_n$ and $\Deh(n) = \qb_n-\qh$.  Note that $\ket{n,k_n}$ is not the eigenbasis of $\H_0(n)$, but rather $\H_0(n)\ket{n,k_n}$ retains the original tridiagonal structure in $n$, i.e.\
\beqn
\H_0(n)\ket{n,k_n} &=& (g_n+n)\ket{n,k_n}-g_n\ket{n+1,k_n}\nonumber\\
&&	-n\ket{n-1,k_n}+\rho k\ket{n,k_n},
\eeqn
where Eqns.\ \ref{b1}, \ref{b3}, \ref{ap}, and \ref{am} are recalled in applying the first term of $\H_0(n)$.

Projecting the conjugate state $\bra{n,k_n}$ onto Eqn.\ \ref{eom_3} yields, after some simplification (cf.\ Appendix C), the equation of motion
\beqn
\label{nkn_final}
&&g_{n-1}G_{n-1,k}+(n+1)G_{n+1,k}%\nonumber\\
	-(g_n+n+\rho k)G_{nk}\nonumber\\
&=&	-g_{n-1}\sum_{\ell=1}^kV^-_{n\ell}G_{n-1,k-\ell}%\nonumber\\
	-(n+1)\sum_{\ell=1}^kV^+_{n\ell}G_{n+1,k-\ell}\nonumber\\
&&	+\rho\Delta_nG_{n,k-1},
\eeqn
where
\beqn
\Delta_n &=& \qb_n-q_n,\\
\label{Vnl}
V^\pm_{n\ell} &=& \frac{(-Q^\pm_n)^\ell}{\ell!},
\eeqn
and $Q^\pm_n = \qb_n-\qb_{n\pm1}$.
Eqn.\ \ref{nkn_final} can be written linear algebraically as
\beqn
\label{nkn_mat}
\vec{G}_k&=&\left({\bf T}^k\right)^{-1}
	\left\{ ({\bf S}^-\vec{g})*{\rm diag}[{\bf V}^-({\bf S}^-\tilde{{\bf G}})^T]\right.\nonumber\\
&&	\left.+({\bf S}^+\vec{n})*{\rm diag}[{\bf V}^+({\bf S}^+\tilde{{\bf G}})^T]\right.\nonumber\\
&&	\left.+\rho \vec{\Delta}*\vec{G}_{k-1} \right\},
\eeqn
where ${\bf X}^T$ indicates the transpose of ${\bf X}$, $*$ denotes an element-by-element product, $S^\pm_{nn'}=\delta_{n\pm1,n'}$ are super- ($+$) and subdiagonal ($-$) matrices, $T^k_{nn'}=g_{n-1}\delta_{n-1,n'}+(n+1)\delta_{n+1,n'}-(g_n+n+\rho k)\delta_{nn'}$ is a tridiagonal matrix, and $\tilde{{\bf G}}$, which is ${\bf G}$ with the columns reversed (i.e.\ $\tilde{G}_{n\ell} = G_{n,k-\ell}$), is built incrementally in $k$.  As with the $\ket{j,k}$ and $\ket{j,k_j}$ bases, the $k$th term is slaved to the previous $k'<k$ terms; in total the solution requires $K$ inversions of an $N$-by-$N$ matrix.  However, here the task of inversion is simplified because the matrix to be inverted is tridiagonal.  In fact, using the Thomas algorithm \cite{thomas1949elliptic}, we obtain an analytic solution for the case of constant production of the first gene and threshold regulation of the second gene, as described in section \ref{analytic}.

The solution is initialized at $k=0$ using
\beqn
G_{n0}&=&\ip{n,(k=0)_n}{G}=\sum_{n'm}\ip{n}{n'}\ip{0_n}{m}p_{n'm}\nonumber\\
	&=&p_n
\eeqn
(cf.\ Eqn.\ \ref{0_on_n}), where $p_n$ is known (cf.\ Eqn.\ \ref{pn}), and the joint distribution is retrieved via the inverse transform
\beq
\label{retrieve3}
p_{nm}=\ip{n,m}{G}=\sum_k G_{nk}\ip{m}{k_n}.
\eeq

\subsubsection{The $\ket{j,k_n}$ basis}
We now consider a basis which employs both the constant-rate eigenfunctions $\ket{j}$ in the $n$ sector and the $n$-dependent eigenfunctions $\ket{k_n}$ in the $m$ sector.  Expressing the joint distribution directly in terms of the eigenbasis expansion of the generating function
\footnote{Because this basis is a function of three coordinates ($j$, $k$, and $n$), we cannot abstractly define the generating function as in Eqn.\ \ref{Gp}; we must instead work directly with the joint probability distribution.},
we write
\beq
\label{pnm4}
p_{nm} = \sum_{jk}G_{jk}\ip{n,m}{j,k_n},
\eeq
where, as in the $\ket{j,k}$ and $\ket{j,k_j}$ bases, the $\ket{j}$ are parameterized by the constant rate $\gb$, and, as in the $\ket{n,k_n}$ basis, the $\ket{k_n}$ are parameterized by the arbitrary function $\qb_n$.  The inverse of Eqn.\ \ref{pnm4} is
\beq
\label{G4}
G_{jk} = \sum_{nm}p_{nm}\ip{j,k_n}{n,m}.
\eeq

Substituting Eqn.\ \ref{pnm4} into Eqn.\ \ref{mastern} at steady state gives, after some simplification (cf.\ Appendix C),
\beqn
0&=&\sum_{jk}G_{jk}\bra{n,m}\{-\H\ket{j,k_n}\nonumber\\
\label{jkn_temp2}
&&	+\ad_n\gh\ket{j,\delta_-k_n}+\al_n\ket{j,\delta_+k_n}\},
\eeqn
where $\H$ is defined as in Eqn.\ \ref{H}, $\ad_n$ and $\al_n$ act as in Eqns.\ \ref{ap}-\ref{ama},
and
\beq
\ket{\delta_\pm k_n} = \ket{k_{n\pm1}}-\ket{k_n}.
\eeq
We may now partition $\H = \H_0+\H_1$ as
\beqn
\label{H0_4}
\H_0(n)&=&\bd_n\bbar_n+\rho\bd_m\bbar_m(n),\\
\label{H1_4}
\H_1(n)&=&\bd_n\Gah+\rho\bd_m\Deh(n),
\eeqn
with $\bbar_n = \al_n - \gb$ and $\Gah = \gb-\gh$ as in the $\ket{j,k}$ and $\ket{j,k_j}$ bases, and $\bbar_m(n) =  \al_m - \qb_n$ and $\Deh(n) = \qb_n-\qh$ as in the $\ket{n,k_n}$ basis.  Noting that $\ket{j,k_n}$ is the eigenbasis of $\H_0(n)$, i.e.
\beq
\label{jknHeig}
\H_0(n)\ket{j,k_n} = (j+\rho k)\ket{j,k_n},
\eeq
Eqn.\ \ref{jkn_temp2} becomes, after more simplification (cf.\ Appendix C),
\beqn
0&=&-(j+\rho k)G_{j'k}%\nonumber\\
	-\sum_{j'}\Gamma_{j-1,j'}G_{j'k}%\nonumber\\
	-\rho\sum_{j'}\Delta_{jj'}G_{j',k-1}\nonumber\\
\label{jkn_final}
&&	+\sum_\pm\sum_{j'}\sum_{\ell=1}^{k}\Lambda^{\pm\ell}_{jj'}G_{j',k-\ell},
\eeqn
where $\Gamma_{jj'}$ is as in Eqn.\ \ref{Gammajj},
\beqn
\Delta_{jj'}&=&\sum_n\ip{j}{n} (\qb_n-q_n) \ip{n}{j'},\\
\Lambda^{+\ell}_{jj'}&=&\sum_n\ip{j}{n} (n+1) \ip{n+1}{j'} V^+_{nl},\\
\Lambda^{-\ell}_{jj'}&=&\sum_n\ip{j}{n} g_{n-1} \ip{n-1}{j'} V^-_{nl},
\eeqn
and $V^\pm_{n\ell}$ is as in Eqn.\ \ref{Vnl}.
Linear algebraically,
\beqn
\vec{G}_k &=& \left({\bf D}^k+{\bf S}^-{\bf \Gamma}\right)^{-1}\nonumber\\
\label{jkn_mat}
&&	\times\left(-\rho{\bf \Delta}\vec{G}_{k-1}
	+\sum_\pm\sum_{\ell=1}^{k}{\bf \Lambda}^{\pm\ell}\vec{G}_{k-\ell}\right),
\eeqn
with $D^k_{jj'}$ and $S^-_{jj'}$ defined as before (cf.\ Eqn.\ \ref{jk_mat}), revealing once again a lower-triangular equation (i.e.\ each $k$th term is slaved to the previous $k'<k$ terms) requiring only matrix multiplication and the inversion of
a $J$-by-$J$ matrix $K$ times.  Recalling Eqn.\ \ref{G4}, the scheme is initialized using
\beq
G_{j0} = \sum_{nm}p_{nm}\ip{j}{n}\ip{(k=0)_n}{m}=\sum_np_n\ip{j}{n}
\eeq
(cf.\ Eqn.\ \ref{0_on_n}) with known $p_n$ (cf.\ Eqn.\ \ref{pn}), and the joint distribution is retrieved using Eqn.\ \ref{pnm4}.

\subsection{Comparison of the representations}
\label{compare}

The spectral representations in Sec.\ \ref{bases} produce equations of motion with similar levels of numerical complexity. In all cases, the the original two-dimensional master equation has been reduced by the lower triangular structure in the second gene's eigenmode number $k$ to a hierarchy of evaluations of one dimensional problems.  The bases differ in the rate parameters, or equivalently gauge freedoms, that one is free to choose: the $\ket{j,k}$ basis requires two constants $\gb$ and $\qb$; the $\ket{j,k_j}$ basis requires $\gb$ and a $J$-valued vector $\qb_j$; the $\ket{n,k_n}$ basis requires a $N$-valued vector $\qb_n$; and the $\ket{j,k_n}$ basis requires $\gb$ and $\qb_n$.

The bases also differ in the types of problems for which they are most suitable.  For example, the $\ket{j,k}$, $\ket{j,k_j}$, and $\ket{j,k_n}$ bases, which all expand the parent species in eigenfunctions $\ket{j}$, are best when a cutoff in $j$ is most appropriate, such as when the parent distribution is a Poisson.  The $\ket{n,k_n}$ basis, on the other hand, is useful when a cutoff is $n$ is most appropriate, such as when the parent species is concentrated at low protein number.  Different bases are more robust to numerical errors for different regulation functions as well: the $\ket{n,k_n}$ and $\ket{j,k_n}$ bases, which both rely upon repeated manipulation of the object $Q^\pm_n = \qb_n-\qb_{n\pm1}$, are best for smooth regulation functions, for which the differences between $\qb_n$ and $\qb_{n+1}$ are small; the $\ket{j,k}$ and $\ket{j,k_j}$ bases on the other hand, which involve the deviations $\qb-q_n$ and $\qb_j-q_n$ respectively, are less susceptible to numerical error given sharp regulation functions, such as a threshold.

As indicated in Figure \ref{diagram},
the $\ket{j,k}$ basis can be viewed as a special case of either the $\ket{j,k_j}$ basis with $\qb_j=\qb$ (in which case Eqn.\ \ref{jkj_final} reduces to Eqn.\ \ref{jk_final}) or of the $\ket{j,k_n}$ basis with $\qb_n=\qb$ (in which case Eqn.\ \ref{jkn_final} reduces to Eqn.\ \ref{jk_final}).
Although possible in principle, expanding in the $\ket{j,m}$ basis 
does not exploit the natural structure of the problem, 
since it neither retains the tridiagonal structure in $n$ nor gains the lower triangular structure in $k$. This example explicitly shows that not all bases are good candidates for simplifying the master equation.

\begin{figure}
\begin{center}
\includegraphics[scale=0.45]{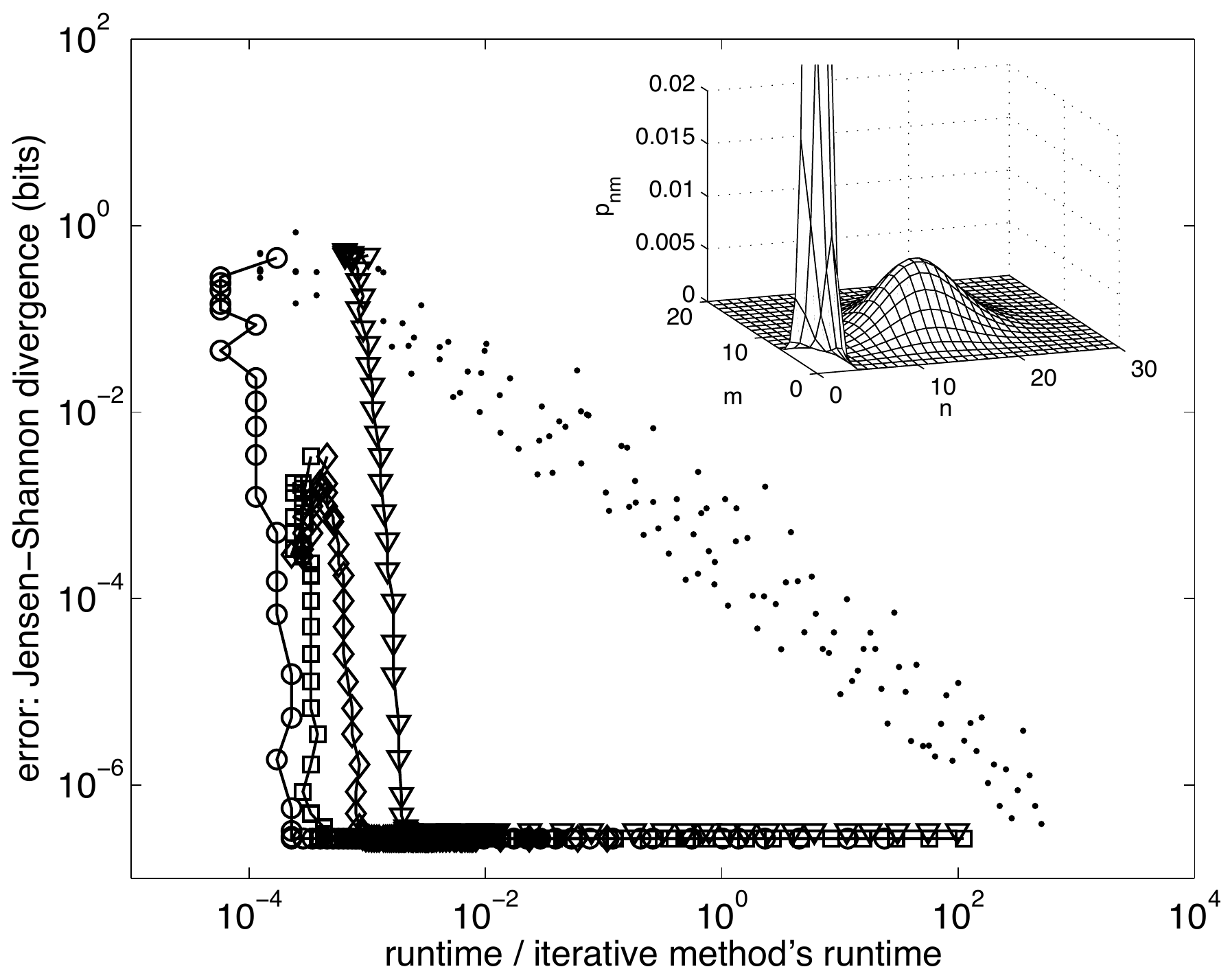}
\linespread{1}
\caption{Error vs.\ runtime for the spectral method and stochastic simulation.  Error is the Jensen-Shannon divergence \cite{Lin} between $p_{nm}$ obtained using the $\ket{n,m}$ basis (via iterative solution of the original master equation) and that obtained using the $\ket{j,k}$ basis (circles; cf.\ Eqn.\ \ref{jk_mat}), the $\ket{j,k_j}$ basis (triangles; cf.\ Eqn.\ \ref{jkj_mat}), the $\ket{n,k_n}$ basis (squares; cf.\ Eqn.\ \ref{nkn_mat}), the $\ket{j,k_n}$ basis (diamonds; cf.\ Eqn.\ \ref{jkn_mat}), or stochastic simulation \cite{Gillespie} (dots).  Runtimes are scaled by that of the iterative solution, $150$ seconds (in MATLAB).  Spectral basis data is obtained by varying $K$, the cutoff in the eigenmode number $k$ of the second gene; simulation data is obtained by varying the integration time.  The input distribution $p_n=\pi_1e^{-\lambda_1}\lambda_1^n/n!+(1-\pi_1)e^{-\lambda_2}\lambda_2^n/n!$ (from which $g_n$ is calculated via Eqn.\ \ref{gn}) is a mixture of two Poisson distributions with $\lambda_1=0.5$, $\lambda_2=15$, and $\pi_1=0.5$.  The regulation function $q_n=q_-+(q_+-q_-)n^\nu/(n^\nu+n_0^\nu)$ is a Hill function with $q_-=1$, $q_+=11$, $n_0=7$, and $\nu=2$.  The gauge choices used (cf.\ Fig.\ \ref{diagram}) are $\gb=\sum_np_ng_n$, $\qb=\sum_np_nq_n$, $\qb_n=q_n$, and $\qb_j=\sum_n\ip{j}{n}q_n\ip{n}{j}$.  The cutoffs used are $J=80$ for the eigenmode number $j$ of the first gene and $N=50$ for the protein numbers $n$ and $m$.  {\bf Inset:} The joint probability distribution $p_{nm}$. The peak at low protein number extends to $p_{00}\approx0.1$.}
\label{fast}
\end{center}
\end{figure}

The strength of all the spectral bases discussed in this section, and of the proposed spectral method in general, is that it allows for a fast and accurate calculation of full steady state probability distributions of the number of protein molecules in a gene regulatory network. In Fig.\ \ref{fast} we demonstrate this property for the two-gene system by plotting error versus computational runtime for each spectral basis, as well as for a stochastic simulation using a varying step Monte Carlo procedure \cite{Gillespie}.  For error we use the Jensen-Shannon divergence \cite{Lin} (a measure in bits between two probability distributions) between the distribution $p_{nm}$ computed in the $\ket{n,m}$ basis (via iterative solution of the original master equation) and the distribution computed either via the spectral formulae in this section or by stochastic simulation.  We plot this measure against the runtime of each method, scaled by the runtime of the iterative solution in the $\ket{n,m}$ basis (all numerical experiments are performed in MATLAB).  We find that the computations via the spectral bases achieve accuracy up to machine precision $\sim$$10^3$-$10^4$ times faster than the iterative method's runtime and $\sim$$10^7$-$10^8$ times faster than the runtime necessary for the stochastic simulation to achieve the same accuracy.  Computation in the $\ket{j,k}$ basis is most efficient since its equation of motion is simplest (cf.\ Eqn.\ \ref{jk_mat}); the $\ket{j,k_j}$ and $\ket{j,k_n}$ bases tend to be slightly less efficient since they require inner loops over $\ell$ (cf.\ Eqns.\ \ref{jkj_mat} and \ref{jkn_mat}). Fig.\ \ref{fast} demonstrates the tremendous gain in performance achieved by the joint analytic-numerical spectral method over traditional simulation approaches.

\subsection{An analytic solution}
\label{analytic}

In general, the equations of motion in the spectral representations (Eqns.\ \ref{jk_final}, \ref{jkj_final}, \ref{nkn_final}, and \ref{jkn_final}) need to be evaluated numerically. In the case of the $\ket{n,k_n}$ basis, however, we can exploit the tridiagonal structure of Eqn.\ \ref{nkn_final} to find an exact analytic solution. Specifically, in the case of a Poisson parent ($g_n=g=$ constant) and for threshold regulation, i.e.\
\beq
\label{thresh}
q_n=
\begin{cases}
q_- & {\rm for\ } n\le n_0 \\
q_+ & {\rm for\ } n>n_0,
\end{cases}
\eeq
setting $\qb_n=q_n$ makes Eqn. \ref{nkn_final}
\beqn
\label{trisimp}
&&gG_{n-1,k}+(n+1)G_{n+1,k}-(\rho k+g+n)G_{nk}\nonumber\\
&&=-g\phi^-_k\delta_{nn_1}-n_1\phi^+_k\delta_{nn_0},
\eeqn
where
\beqn
\phi^-_k&=&\sum_{\ell=1}^k\frac{(-\Delta)^\ell}{\ell!}G_{n_0,k-\ell},\\
\phi_k^+&=&\sum_{\ell=1}^k\frac{\Delta^\ell}{\ell!}G_{n_1,k-\ell},
\eeqn
$\Delta=q_+-q_-$, and $n_1=n_0+1$.  Eqn.\ \ref{trisimp} is solved using the tridiagonal matrix algorithm (also called the Thomas algorithm \cite{thomas1949elliptic}), as described in detail in Appendix D.  The result is an analytic expression for the $k$th column of $G_{nk}$ in terms of its previous columns (i.e.\ the matrix inversion has been performed explicitly),
\beqn
G_{nk} &=& \frac{n_1}{g}\frac{(n_0-1)!}{n!}\frac{\eta_n^k}{\eta_{n_0-1}^k}\nonumber\\
\label{Gana}
&&	\times
\begin{cases}
	\left(\phi_k^++f_kF_{n_1}^k\right)/(\epsilon_{n_0}^k-1) & n\le n_0 \\
	f_kF_n^k/\prod_{i=n_0}^{n-1}(\epsilon_i^k-1) & n>n_0,
\end{cases}\qquad
\eeqn
where
\beqn
f_k &=& \phi^+_k-\frac{gn_0}{n_1}\frac{\eta^k_{n_0-1}}{\eta^k_{n_0}}(\epsilon^k_{n_0}-1)\phi_k^-,\\
F_n^k &=& \sum_{i=0}^{N-n}\prod_{\ell=n}^{n+i}\frac{1}{\epsilon_\ell^k-1},\\
\epsilon_n^k &=& \frac{\rho k+g+n}{gn}\frac{\eta^k_n}{\eta^k_{n-1}},\\
\eta_n^k &=& \sum_{i=0}^n \frac{n!}{i!(n-i)!}g^{n-i} \prod_{\ell=0}^{i-1} (\rho k+\ell)\\
\label{etaint}
&=& \frac{1}{\Gamma(\rho k)}\int_0^t dt \e{-t} t^{(\rho k-1)}
(g+t)^n,
\eeqn
and $N$ is the cutoff in protein number $n$.  Along with the analytic form of the mixed product
\beqn
\ip{m}{k_n} &=& (-1)^ke^{-q_n}q_n^mk!\nonumber\\
&&\times\sum_{\ell=0}^{\min(m,k)}\frac{1}{\ell!(m-\ell)!(k-\ell)!(-q_n)^\ell}\qquad
\eeqn
(cf.\ Appendix A), Eqn.\ \ref{Gana} in the limit $N\rightarrow\infty$ constitutes an exact analytic solution for the joint distribution $p_{nm}$, as calculated using Eqn.\ \ref{retrieve3}.

\subsection{The threshold-regulated gene approximates the on/off gene}
\label{threshonoff}

If a gene is regulated via a threshold function (cf.\ Eqn.\ \ref{thresh}), its steady state protein distribution $p_m$ can be well approximated by the two-state process discussed in Sec.\ \ref{onoff}.  To make the connection clear, we first observe that the off-state ($z=-$) corresponds to the first gene expressing the same or fewer proteins $n$ than the threshold $n_0$, i.e.\
\beq
\label{pm-}
p_m^-=\sum_{n\le n_0}p_{nm},
\eeq
and the on-state ($z=+$) corresponds to the first gene expressing more proteins than the threshold, i.e.\
\beq
\label{pm+}
p_m^+=\sum_{n>n_0}p_{nm}.
\eeq
The dynamics of $p_m^\pm$ are then obtained by summing the master equation for two-gene regulation, Eqn.\ \ref{mastern}, over either all $n\le n_0$ or all $n>n_0$, giving
\beqn
\label{burstMEa}
\dot{p}_m^\pm &=& \rho\left[q_\pm p_{m-1}^\pm+(m+1)p_{m+1}^\pm-(q_\pm+m)p_m^\pm\right]\nonumber\\
\label{threshsum1}
&&	\mp n_1p_{n_1m}\pm g_{n_0}p_{n_0m},
\eeqn
where $n_1=n_0+1$.  Making the approximations
\beqn
\label{approx1}
\frac{p_m^-}{\pi_-} = p(m|-) &\approx& p(m|n_0) = \frac{p_{n_0m}}{p_{n_0}}, \\
\label{approx2}
\frac{p_m^+}{\pi_+} = p(m|+) &\approx& p(m|n_1) = \frac{p_{n_1m}}{p_{n_1}},
\eeqn
where
\beqn
\pi_-&=&\sum_mp_m^-=\sum_{n\le n_0}p_n,\\
\pi_+&=&\sum_mp_m^+=\sum_{n>n_0}p_n
\eeqn
are the total probabilities of being in the off- and on-states respectively, and noting from Eqn.\ \ref{gn} that $g_{n_0}=n_1p_{n_1}/p_{n_0}$, Eqn.\ \ref{burstMEa} at steady state becomes
\beqn
\label{burstMEb}
0 &=& q_zp_{m-1}^z+(m+1)p_{m+1}^z-(q_z+m)p_m^z \nonumber\\
&&	+\sum_{z'}\Omega_{zz'} p_m^{z'},
\eeqn
with $z=\pm$ and
\beq
\label{Omega2a}
{\bf \Omega} = \begin{pmatrix} -\omega_+ & \omega_- \\ \omega_+ & -\omega_- \end{pmatrix},
\eeq
where
\beq
\label{omeff}
\omega_\pm = \frac{n_1p_{n_1}}{\rho\pi_\mp}.
\eeq
Eqns.\ \ref{burstMEb}-\ref{Omega2a} have the same form as Eqns.\ \ref{burstME} and \ref{Omega2} at steady state with $n\rightarrow m$ and $g\rightarrow q$, and Eqn.\ \ref{omeff} relates the effective switching rates $\omega_\pm$ to input and regulation parameters $p_{n_1}$, $\pi_\pm$, and $n_1$, and the ratio $\rho$ of the degradation rate of the second gene to that of the first.  Note that Eqn.\ \ref{omeff} satisfies
\beq
\frac{\pi_-}{\pi_+} = \frac{\omega_-}{\omega_+},
\eeq
in agreement with Eqn.\ \ref{piom}, and exhibits the intuitive behavior that increasing $\rho$ (i.e. decreasing the relative response rate of the first gene) is equivalent to decreasing the switching rates $\omega_\pm$.

\begin{figure}
\begin{center}
\includegraphics[scale=0.65]{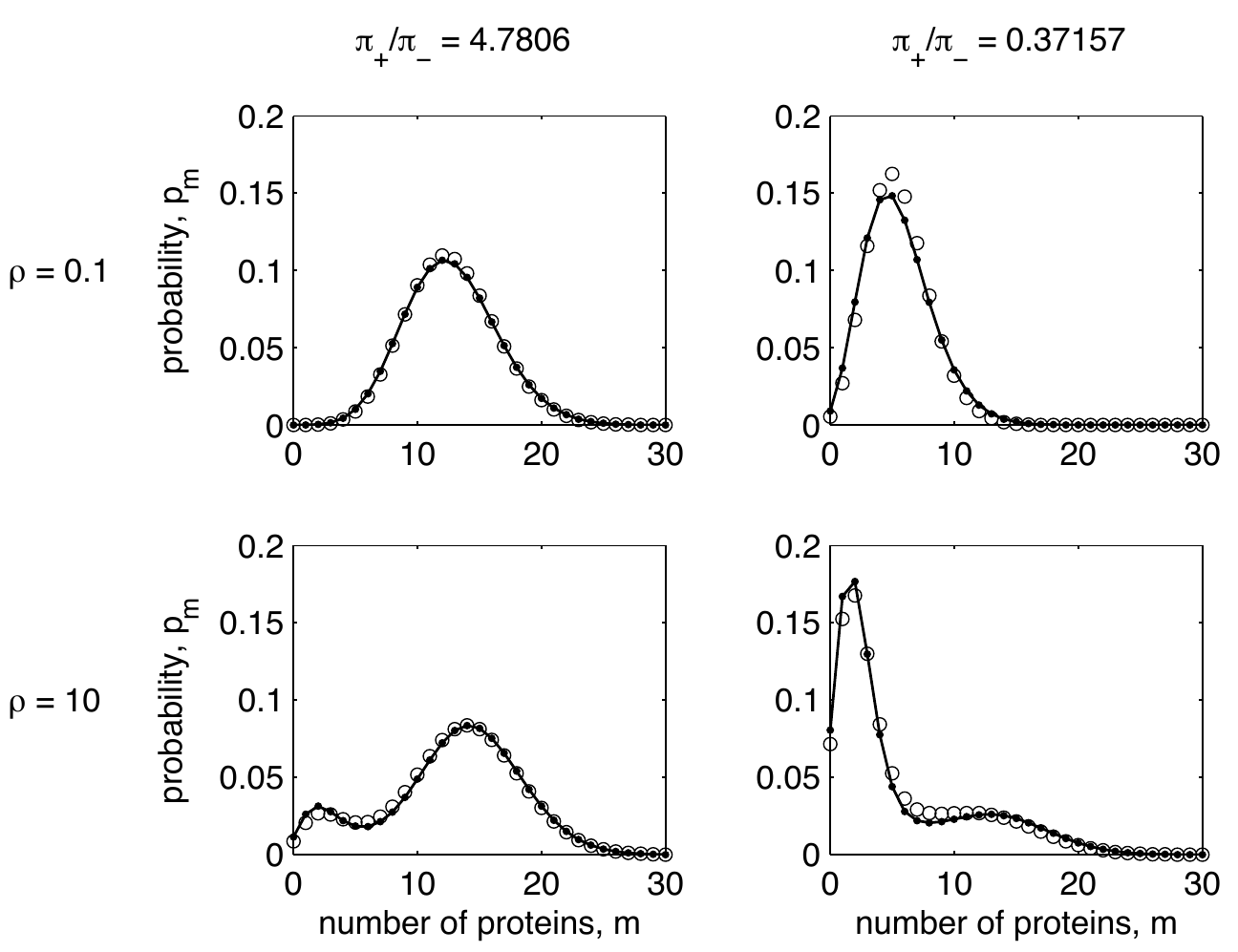}
\linespread{1}
\caption{Protein distributions for a gene regulated by a threshold function (dots; calculated via Eqn.\ \ref{jk_final}) and a gene with two stochastic states (circles; calculated via Eqn.\ \ref{cEoMbar2}).  The relationship between regulation parameters and state transition rates is given by Eqn.\ \ref{omeff}.  In all panels the input to the regulation is a Poisson distribution with mean $g=7$, and the regulation rates (cf.\ Eqn.\ \ref{thresh}) are $q_-=2$ and $q_+=15$.  In the first column the threshold is $n_0=4$ making $\pi_+=0.827>\pi_-=0.173$; in the second column $n_0=8$ making $\pi_+=0.271<\pi_-=0.729$.  In the first row the ratio of the second gene's degradation rate to that of the first is $\rho=0.1$; in the second row $\rho=10$.}
\label{b2t}
\end{center}
\end{figure}

A comparison of the distributions of a threshold-regulated gene with those of an on/off gene for various parameter settings reveals that Eqns.\ \ref{approx1}-\ref{approx2} are a good approximation.  Fig.\ \ref{b2t} shows a demonstration for a threshold-regulated system with a Poisson input distribution.  In the first column, the mean $g$ of the input lies above the threshold $n_0$, making the output more likely to be in the on-state, i.e. $\pi_+>\pi_-$; in the second column, $g<n_0$, making $\pi_+<\pi_-$.  In the first row $\rho<1$; in the second row $\rho>1$, corresponding to lower effective switching rates $\omega_\pm$ and producing bimodal distributions with peaks near the on/off rates $q_\pm$.  In all examples, the approximation as a two-state process with switching rates given by Eqn.\ \ref{omeff} agrees well with the actual output from threshold regulation.

\section{Regulation with bursts}
\label{burstreg}

The final system we consider combines the multi-state process used to model bursts of expression in Sec.\ \ref{burst} with gene regulation as discussed in Sec.\ \ref{reg}.  Specifically we consider a system of two species, with protein numbers $n$ and $m$, existing in $Z$ possible states, distinguished by the settings of the two production rates $g_z$ and $q_z$ respectively, where $1\le z\le Z$.  Regulation is achieved by allowing the rates of transition among states affecting the production of the second gene to depend on the number $n$ of proteins expressed by the first gene.  Recalling Eqns.\ \ref{burstME} and \ref{mastern}, the master equation describing the evolution of the joint probability distribution $p_{nm}^z$ reads
\beqn
\label{brme}
\dot{p}_{nm}^z &=& g_zp^z_{n-1,m}+(n+1)p^z_{n+1,m}-(g_z+n)p^z_{nm}\nonumber\\
&&	+\rho\left[q_zp^z_{n,m-1}+(m+1)p^z_{n,m+1}-(q_z+m)p^z_{nm}\right]\nonumber\\
&&	+\sum_{z'}\Omega_{zz'}(n) p_{nm}^{z'},
\eeqn
where the dependence of the stochastic matrix $\Omega_{zz'}$ on $n$ incorporates the regulation.

As with the previously discussed models, Eqn.\ \ref{brme} benefits from spectral expansion, and for simplicity we present only the formulation in the $\ket{j,k}$ basis, parameterized by constant rates $\gb$ and $\qb$ respectively, as in Secs.\ \ref{specz} and \ref{jkbasis}.  As before the first step is to define the generating function
\beq
\label{Gnmz}
\ket{G_z} = \sum_{nm}p_{nm}^z\ket{n,m},
\eeq
with which Eqn.\ \ref{brme}, upon summing over $n$ and $m$ against $\ket{n,m}$, becomes
\beq
\label{brme2}
\ket{\dot{G}_z} = -\H_z\ket{G_z}+\sum_{z'}\Om_{zz'}\ket{G_{z'}},
\eeq
where
\beqn
\H_z &=& \bd_n\bl_{nz}+\rho\bd_m\bl_{mz}\\
\label{b1a}
\bd_n &=& \ad_n-1,\\
\label{b2a}
\bd_m &=& \ad_m-1,\\
\label{b3a}
\bl_{nz} &=& \al_n - g_z,\\
\label{b4a}
\bl_{mz} &=& \al_m - q_z,
\eeqn
and $\Om_{zz'}$ is $\Omega_{zz'}(n)$ with every instance of $n$ replaced by the number operator $\ad_n\al_n$.  Defining $\bbar_n = \al_n-\gb$ and $\bbar_m=\al_m-\qb$, we partition the Hamiltonian as $\H_z = \H_0+\H_1^z$, with
\beq
\H_0 = \bd_n\bbar_n+\rho\bd_m\bbar_m
\eeq
the operator of which $\ket{j,k}$ is the eigenbasis, i.e.\
\beq
\H_0\ket{j,k} = (j+\rho k)\ket{j,k},
\eeq
and
\beq
\H_1^z = \bd_n\Gamma_z+\rho\bd_m\Delta_z
\eeq
capturing the deviations $\Gamma_z=\gb-g_z$ and $\Delta_z = \qb-q_z$ of the constant rates from the state-dependent rates.  Upon expanding the generating function in the eigenbasis,
\beq
\label{Gjkz}
\ket{G_z} = \sum_{jk}G^z_{jk}\ket{j,k},
\eeq
and taking dummy indices $j\rightarrow j'$ and $k\rightarrow k'$, projecting the conjugate state $\bra{j,k}$ onto Eqn.\ \ref{brme2} gives
\beqn
\label{brme3}
\dot{G}^z_{jk} &=& -(j+\rho k)G^z_{jk}-\Gamma_zG^z_{j-1,k}-\Delta_zG^z_{j,k-1}\nonumber\\
&&	+\sum_{z'}\sum_{j'}\bra{j}\Om_{zz'}\ket{j'}G^{z'}_{j'k},
\eeqn
where the components of $\Om_{zz'}$ need only be evaluated in the $j$ sector, not the $k$ sector, because the transition rates depend on only $n$, not $m$ (cf.\ Eqn.\ \ref{brme}).  Like Eqns.\ \ref{cEoMbar2} and \ref{jk_final}, Eqn.\ \ref{brme3} is subdiagonal in $k$ and thus far more efficient to solve than the original master equation, Eqn.\ \ref{brme}, as we demonstrate for a special case in the next section.  The joint distribution is retrieved from $G^z_{jk}$ via inverse transform,
\beq
\label{pnmzret}
p^z_{nm} = \sum_{jk}\ip{n}{j}G^z_{jk}\ip{m}{k},
\eeq
with the mixed products calculated as in Appendix A.

\subsection{The four-state process}

As a simple example of the model in Eqn.\ \ref{brme}, we consider a system in which each of the two species has an on-state and an off-state, and the transition rate of the second species to its on-state is a function of the number of copies of the first species.  This system models both (i) a single gene for which the production of proteins depends on the number of transcripts, and each is produced in on- and off-states by the binding and unbinding of ribosomes and RNA polymerase respectively, and (ii) one gene regulating another with each undergoing burst-like expression.

There are a total of $Z=4$ states, i.e.\
\beqn
p^z_{nm} &=& (p^{--}_{nm},p^{+-}_{nm},p^{-+}_{nm},p^{++}_{nm}).
\eeqn
where the first signed index denotes the state of the first gene (with protein count $n$) and the second signed index denotes the state of the second gene (with protein count $m$). Defining $g_\pm$ as the production rates of the first species in its on- ($+$) and off-states ($-$), and similarly $q_\pm$ for the second species, the production rates of the $Z=4$ states are:\
\beqn
g_z &=& (g_-,g_+,g_-,g_+),\\
q_z &=& (q_-,q_-,q_+,q_+).
\eeqn
Defining $\omega_\pm$ as the transition rates of the first species to ($+$) and from ($-$) its on-state, and similarly $\alpha_\pm$ for the second species, the transition matrix takes the form

\medskip
$\Omega_{zz'}(n) = $
\beqn
\begin{pmatrix}
	-\omega_+-\alpha_+(n) & \omega_- & \alpha_- & 0 \\
	\omega_+ & -\omega_--\alpha_+(n) & 0 & \alpha_- \\
	\alpha_+(n) & 0& -\omega_+-\alpha_- & \omega_- \\
	0 & \alpha_+(n) & \omega_+ & -\omega_--\alpha_- \\
\end{pmatrix}.\nonumber
\eeqn
\beq
\eeq
The simple form $\alpha_+(n) = cn^\nu$ for constant $c$ and integer $\nu$ corresponds to the first species activating the second as a multimer, with $\nu$ the order of the multimerization.
In the limit of fast switching this description reduces to a Hill function with cooperativity $\nu$ \cite{Walczak2}.
Recalling that $\bd_n=\ad_n-1$ and $\bbar_n = \al_n-\gb$, the $n$-dependent terms of $\bra{j}\Om_{zz'}\ket{j'}$ are evaluated as
\beq
\label{Omeval}
\bra{j}\alpha_+(\ad_n\al_n)\ket{j'} = c\bra{j}[(\bd_n+1)(\bbar_n+\gb)]^\nu\ket{j'},
\eeq
Since $\bd_n$ and $\bbar_n$ raise and lower $\ket{j'}$ states respectively (cf.\ Eqns.\ \ref{bp}-\ref{bm}), the modified transition matrix $\bra{j}\Om_{zz'}\ket{j'}$ is nearly diagonal, with nonzero terms only for $|j-j'|\le\nu$.

Eqn.\ \ref{brme3} at steady state,
\beqn
\label{brme4}
&&-(j+\rho k)G^z_{jk}-\Gamma_zG^z_{j-1,k}+\sum_{z'}\sum_{j'}\bra{j}\Om_{zz'}\ket{j'}G^{z'}_{j'k}\nonumber\\
&&=\Delta_zG^z_{j,k-1},
\eeqn
is solved successively in $k$, requiring the inversion of a $4J$-by-$4J$ matrix $K$ times.  It is initialized at $k=0$ by computing the null space of the left hand side and normalizing with $\sum_z G_{00}^z = 1$ (cf.\ Eqn.\ \ref{init2}).  The joint distribution $p_{nm}^z$ is retrieved via inverse transform (Eqn.\ \ref{pnmzret}).

With $\nu=2$, a typical solution of Eqn.\ \ref{brme4} takes a few seconds (in MATLAB), which, depending on the cutoff $N$, is $\sim$$10^2$-$10^3$ times faster than direct solution of the master equation, Eqn.\ \ref{brme}, by iteration, for equivalent accuracy. The advantage of such a large efficiency gain is that it allows repeated evaluations of the governing equation, necessary for parameter inference or optimization \cite{Walczak}.  We demonstrate this possibility in the next section by finding and interpreting the solutions that optimize the information flow from the first to the second species.

\subsection{The information-optimal solution}

Cells use regulatory processes to transmit relevant information from one species to the next \cite{Gomperts, Ting, Detwiler, Bolouri, Bassler}.  Information processing is quantified by the mutual information $I$, which, between the first and second species in the four-state process, is
\beq
\label{I}
I = \sum_{nm}p_{nm}\log_2\frac{p_{nm}}{p_np_m},
\eeq
where the distributions $p_{nm}$, $p_n$, and $p_m$ are obtained from summing the joint distribution $p_{nm}^z$ (cf.\ Eqn.\ \ref{pnmzret}), and the log is taken with base 2 to give $I$ in bits.

\begin{figure}
\begin{center}
\includegraphics[scale=0.43]{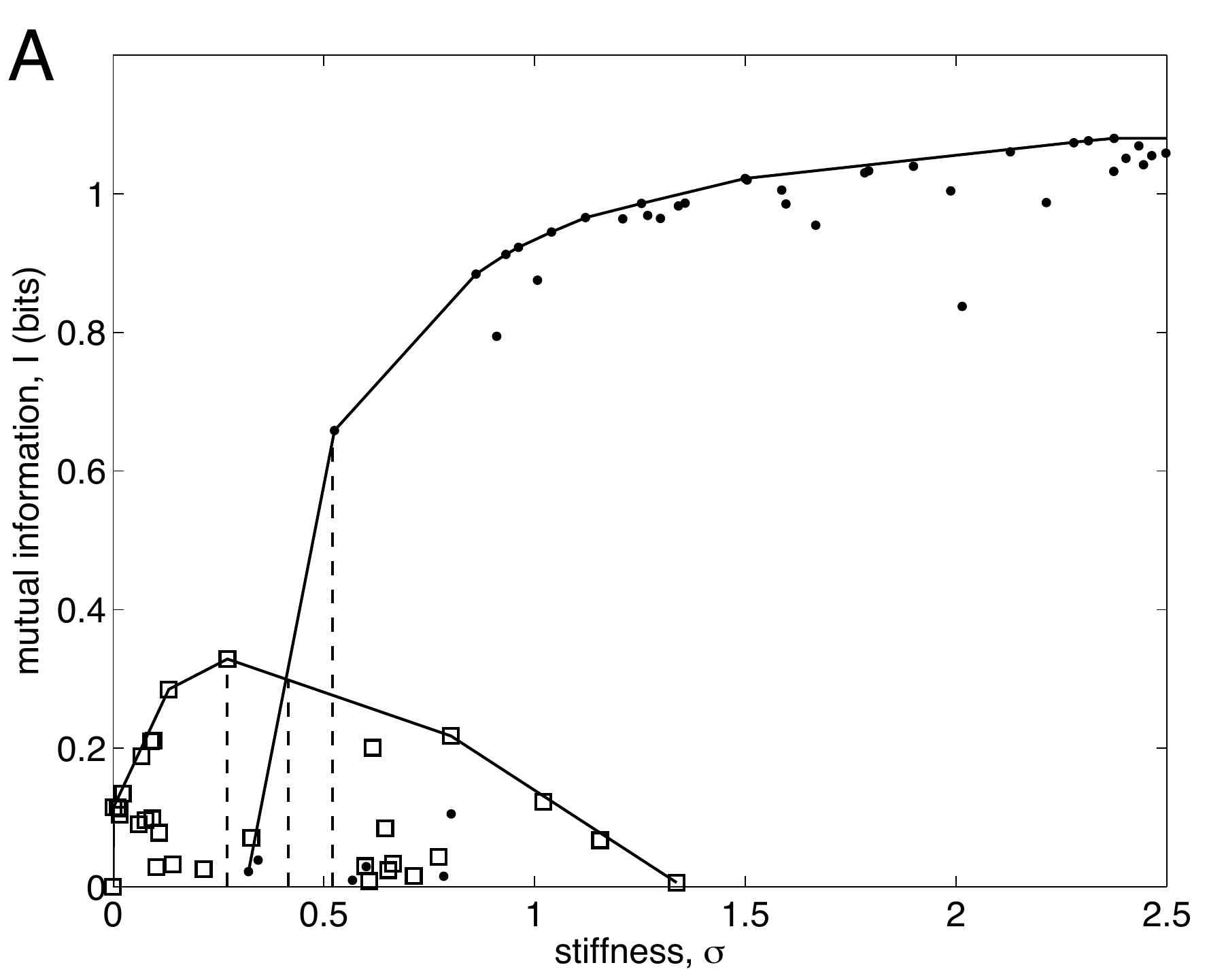}
\includegraphics[scale=0.43]{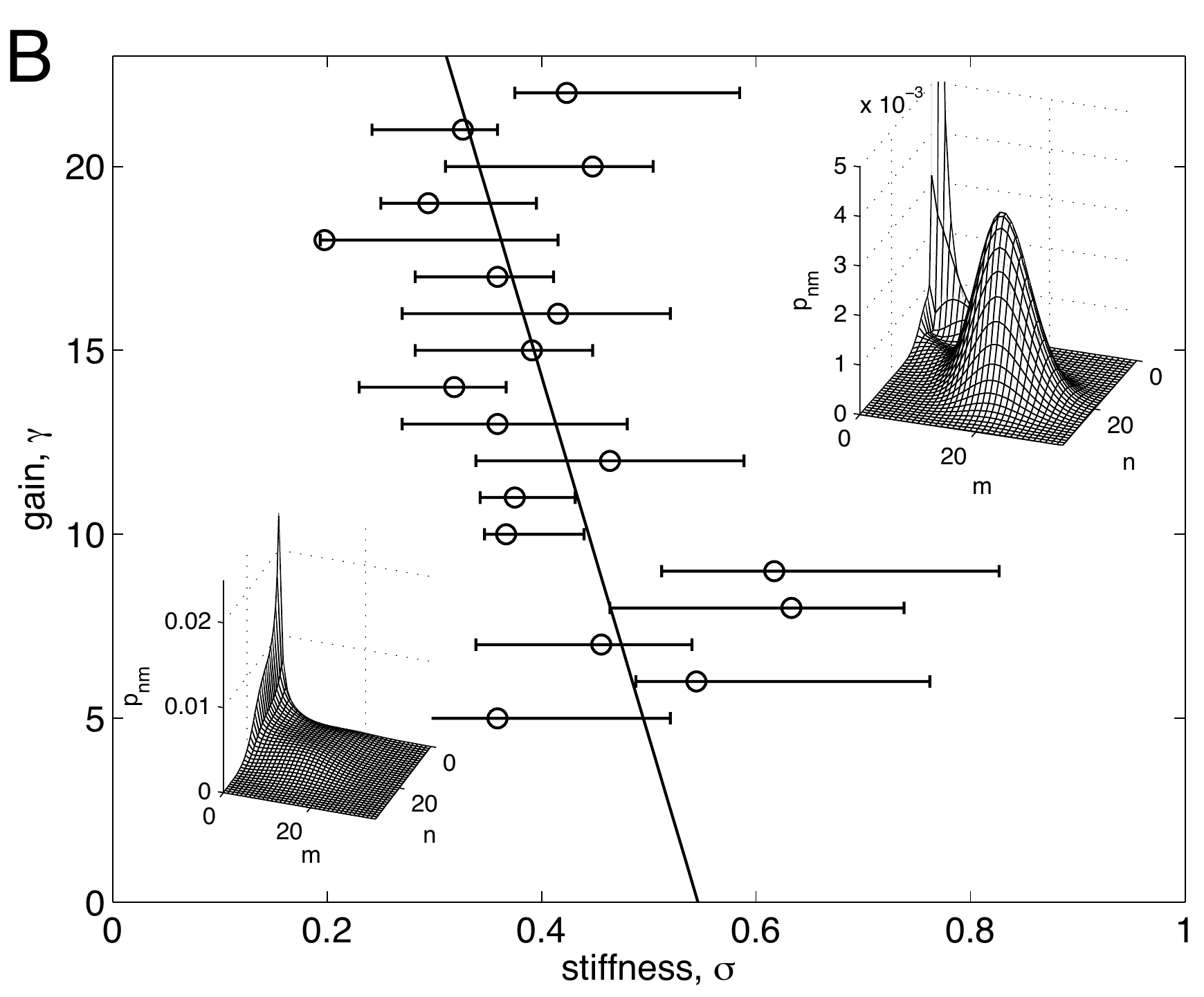}
\linespread{1}
\caption{{\bf A:} Mutual information $I$ (cf.\ Eqn.\ \ref{I}) versus stiffness $\sigma$ (cf.\ Eqn.\ \ref{stiffness}) for fixed gain ($\gamma=16$, cf.\ Eqn.\ \ref{gain}), obtained by optimizing Eqn.\ \ref{L} for $\lambda$ values between $10^{-3}$ and $10^1$.  Squares denote solutions whose joint distribution $p_{nm}$ has one peak (cf.\ B, lower left inset), and dots denote solutions for which $p_{nm}$ has two peaks (cf.\ B, upper right inset).  Solid lines show the convex hulls of the one- and two-peaked solutions.  Dotted lines indicate the stiffness value at which the hulls intersect and the stiffness values of the hull points to the left and right of the intersection.  {\bf B:} Phase diagram between one- and two-peaked optimal solutions in the gain-stiffness plane.  Circles and left and right error bars at each gain are determined by the stiffness values at the intersection of the one- and two-peak convex hulls and at the hull points to the left and right of the intersection respectively (see dotted lines for the example case in A).  Solid line shows a line of best fit.  Insets show examples of one- (lower left) and two-peaked (upper right) optimal distributions $p_{nm}$.
}
\label{phase}
\end{center}
\end{figure}

Upon optimization of $I$ for the four-state process, two distinct types of optimal solutions become clear: those in which the distribution $p_{nm}$ has one peak, and those in which $p_{nm}$ has two peaks.  The former occur when copy number is constrained to be low, and switching rates are constrained to be near the decay rates of both species, producing a single peak at low copy number (see lower left inset of Fig.\ \ref{phase}B).  As these constraints are lifted, it is optimal for the switching rates of the parent species to become much less than the decay rate.  The slow switching produces a second peak whose location is specified by the on-rate of each species (see upper right inset of Fig.\ \ref{phase}B).

To quantify the transition between the two types of solutions, we numerically optimized mutual information over parameters $g_+, q_+, \omega_-, \omega_+, \alpha_-$, and $c$ (the off-rates $g_-$ and $q_-$ were fixed at $0$; the cooperativity $\nu$ was fixed at $2$; and the decay rate ratio $\rho$ was fixed at $1$).  Information may always be trivially optimized by allowing infinite copy number or arbitrary separation of relevant timescales.  We limit copy number by constraining the gain
\beq
\label{gain}
\gamma = \frac{\Gamma+\Delta}{2},
\eeq
defined as the average of the parent gain $\Gamma = g_+-g_-$ and the child gain $\Delta = q_+-q_-$.  Since $g_-=q_-=0$, the maximum number of particles is dictated by the on-rates $g_+$ and $q_+$, and thus constraining $\gamma$ limits the copy number.  We limit separation between the switching timescales and the decay timescales by constraining the stiffness
\beqn
\label{stiffness}
\sigma &=& \frac{1}{4}\left( |\log_{10}\omega_-|+|\log_{10}\omega_+|\right.\nonumber\\
&&	\left.+|\log_{10}\alpha_-|+|\log_{10}\left[\alpha_+\avg{n^\nu}\right]|\right),
\eeqn
where the average $\avg{n^\nu}$ is taken over $p_n$. Stiffness $\sigma$ is the average of the absolute deviation of (the logs of) all four switching rates from the (unit) decay rates, so constraining $\sigma$ prevents fast or slow switching.  Gain is fixed by varying $g_+$ and $q_+$ such that $\gamma$ is a constant, and stiffness is constrained by optimizing the objective function
\beq
\label{L}
\L = I - \lambda \sigma
\eeq
for a given value of the Lagrange multiplier $\lambda$.

As shown in Fig.\ \ref{phase}A, one-peaked solutions are more informative at low stiffness, while two-peaked solutions are more informative at high stiffness.  We compute the convex hulls of the one- and two-peaked data to remove suboptimal solutions, and the transition occurs at the stiffness value at which the convex hulls intersect (cf.\ Fig.\ \ref{phase}A).  Repeating this procedure for many choices of gain allows one to trace out the phase transition shown Fig.\ \ref{phase}B, which makes clear that one-peaked solutions are most informative at low stiffness, two-peaked solutions are most informative at high stiffness, and the critical stiffness decreases weakly with increasing gain.

\section{Conclusions}

The presented spectral method exploits the linearity of the master equation to solve for probability distributions directly by expanding in the natural eigenfunctions the linear operator.
We demonstrate the method on three models of gene expression: a single gene with multiple expression states, a gene regulatory cascade, and a model that combines multi-state expression with explicit regulation through binding of transcription factor proteins.

The spectral method permits huge computational gains over simulation.  As demonstrated for all spectral expansions of the two-gene cascade (cf.\ Fig.\ \ref{fast}), directly solving for the distribution via the spectral method is $\sim$$10^7$-$10^8$ times faster than building the distribution from samples using a simulation technique. This massive speedup makes possible optimization and inference problems requiring full probability distributions that were not computationally feasible previously.  For example, by optimizing information flow in a two-gene cascade in which both parent and child undergo two-state production, we reveal a transition from a one-peaked to a two-peaked joint probability distribution when constraints on protein number and timescale separation are relaxed.  We emphasize that this optimization would not have been possible without the novel efficiency of the spectral method.

The spectral method also makes explicit the linear algebraic structure underlying the master equation.  In many cases, such as in two-state bursting and the two-gene threshold regulation problem, this leads to analytic solutions.  In general, such as shown in the case of the linear cascade, this leads to a set of natural bases for expansion of the generating function and reveals the features of each basis that are better suited to different types of problems.  Specifically, bases in which the parent species is expanded in eigenfunctions are best when the parent distribution is Poissonian, and bases in which the parent is left in protein number space are best when the parent distribution is concentrated at low protein number.  As well, bases in which the eigenfunctions of the child depend on the number of copies of the parent's protein are best suited for smooth regulation functions, whereas a basis in which the eigenfunctions of the child are parameterized by a constant is more numerically robust for sharp regulation functions such as thresholds.
In all cases the linear algebraic structure of the spectral decomposition yields numerical prescriptions that greatly outperform simulation techniques.
We anticipate that the computational speedup of the method, as well
as the removal of the statistical obstacle of density estimation inherently limiting simulation-based approaches, will make spectral methods such as those demostrated here useful in addressing a wide variety of biological quesitons regarding accurate and efficient modeling of noisy information transmission in biological systems.

\section*{APPENDIX A}

In this appendix we describe two ways to compute the mixed products $\ip{n}{j}$ and $\ip{j}{n}$ between the protein number states $\ket{n}$ and the eigenstates $\ket{j}$: by direct evaluation and by recursive updating.

The direct evaluation follows from Eqns.\ \ref{ip}, \ref{nx}, \ref{jket}, and \ref{cauchy}, and the fact that repeated derivatives of a product follow a binomial expansion.  Introducing $g$ as the rate parameter for the $\ket{j}$ states,
\beqn
\ip{n}{j} &=& \oint \frac{dx}{2\pi i} \ip{n}{x} \ip{x}{j} \\
&=& \oint \frac{dx}{2\pi i} \frac{e^{g(x-1)}(x-1)^j}{x^{n+1}} \\
&=&  \frac{1}{n!} \partial^n_x \left[ e^{g(x-1)}(x-1)^j \right]_{x=0} \\
&=& \frac{1}{n!}\sum_{\ell=0}^n \frac{n!}{\ell! (n-\ell)!}
	\partial^{n-\ell}_x\left[ e^{g(x-1)} \right]_{x=0}\nonumber\\
&&	\times\partial^\ell_x\left[ (x-1)^j \right]_{x=0} \\
&=& \sum_{\ell=0}^n \frac{1}{\ell! (n-\ell)!}\
	\left[ g^{n-\ell}e^{-g} \right]\nonumber\\
&&	\times\left[ \frac{j!}{(j-\ell)!} (-1)^{j-\ell} \theta(j-\ell+1) \right] \\
\label{nj}
&=& (-1)^j e^{-g} g^n j! \xi_{nj},
\eeqn
where
\beq
\label{xi}
\xi_{nj} = \sum_{\ell=0}^{\min(n,j)} \frac{1}{\ell!(n-\ell)!(j-\ell)!(-q)^\ell}.
\eeq
Similarly, noting Eqns.\ \ref{jbra} and \ref{xn},
\beq
\label{jn}
\ip{j}{n} = n!(-g)^j\xi_{nj},
\eeq
with $\xi_{nj}$ as in Eqn.\ \ref{xi}. Eqns.\ \ref{jn} and \ref{nj} clearly reduce to Eqns.\ \ref{0_on_n}-\ref{n_on_0} for the special case $j=0$.

It is more computationally efficient to take advantage of the selection rules in Eqns. \ref{ap}-\ref{ama} and \ref{bp}-\ref{bma} to compute the mixed products recursively.  For example, using Eqns.\ \ref{bp}, \ref{bpdef}, and \ref{apa},
\beqn
\ip{n}{j+1}&=&\bra{n}\bd\ket{j}=\bra{n}(\ad-1)\ket{j}\nonumber\\
\label{dd}
\label{up1}
&=&\ip{n-1}{j}-\ip{n}{j},
\eeqn
which can be initialized using $\ip{n}{0}=e^{-g}g^n/n!$ (cf.\ Eqn.\ \ref{nj}) and updated recursively in $j$. Eqn.\ \ref{dd} makes clear that in $n$ space the $(j+1)$th mode is simply the (negative of the) discrete derivative of the $j$th mode.

Alternatively, Eqns.\ \ref{ama}, \ref{bmdef}, and \ref{bm} give
\beqn
(n+1)\ip{n+1}{j}&=&\bra{n}\al\ket{j}=\bra{n}(\bl+g)\ket{j}\nonumber\\
\label{up2}
&=&j\ip{n}{j-1}+g\ip{n}{j},
\eeqn
which can be initialized using $\ip{0}{j}=(-1)^je^{-g}$ (cf.\ Eqn.\ \ref{nj}) and updated recursively in $n$.

One may similarly derive recursion relations for $\ip{j}{n}$, i.e.\
\beqn
\label{up3}
\ip{j}{n+1}&=&\ip{j-1}{n}+\ip{j}{n},\\
\label{up4}
(j+1)\ip{j+1}{n}&=&n\ip{j}{n-1}-g\ip{j}{n},
\eeqn
initialized with $\ip{j}{0}=(-g)^j/j!$ or $\ip{0}{n}=1$ respectively (cf.\ Eqn.\ \ref{jn}) and updated recursively in $n$ or $j$ respectively.

One may also use the full birth-death operator $\bd\bl$ to derive the recursion relations
\beqn
\label{up5}
(n+1)\ip{n+1}{j}&=&(g+n-j)\ip{n}{j}-g\ip{n-1}{j},\quad\qquad\\
\label{up6}
g\ip{j}{n+1}&=&(g+n-j)\ip{j}{n}-n\ip{j}{n-1},\quad\qquad
\eeqn
initialized with $\ip{0}{j}=(-1)^je^{-g}$ and $\ip{1}{j}=(-1)^je^{-g}(g-j)$ (cf.\ Eqn.\ \ref{nj}), and $\ip{j}{0}=(-g)^j/j!$ and $\ip{j}{1}=(-g)^j(1-j/g)/j!$ (cf.\ Eqn.\ \ref{jn}), respectively, and updated recursively in $n$.  We find Eqns.\ \ref{up5}-\ref{up6} are more numerically stable than Eqns.\ \ref{up1}-\ref{up4}, as the former are two-term recursion relations while the latter are one-term recursion relations.

\section*{APPENDIX B}

In the limit $g_-=0$, Eqn.\ \ref{Ghyp} reads
\beqn
G(x)&=&\frac{\omega_+}{\omega_++\omega_-}e^y
	\Phi[\omega_-,\omega_++\omega_-+1;-y]\nonumber\\
\label{Ghyp2}
&&	+\frac{\omega_-}{\omega_++\omega_-}
	\Phi[\omega_+,\omega_++\omega_-+1;y],
\eeqn
where $y=e^{g_+(x-1)}$.  Using the fact that \cite{Koepf}
\beq
e^y\Phi[\alpha,\beta;-y] = \Phi[\beta-\alpha,\beta;y],
\eeq
Eqn.\ \ref{Ghyp2} can be written
\beqn
G(x)&=&\frac{\omega_+}{\omega_++\omega_-}
	\Phi[\omega_++1,\omega_++\omega_-+1;y]\nonumber\\
&&	+\frac{\omega_-}{\omega_++\omega_-}
	\Phi[\omega_+,\omega_++\omega_-+1;y],
\eeqn
or, noting Eqn. \ref{hyp} and the fact that $\Gamma(z+1)=z\Gamma(z)$ for any $z$,
\beqn
G(x)&=&\sum_j\left(\frac{\omega_+}{\omega_++\omega_-}\frac{\Gamma(j+\omega_++1)}{\Gamma(\omega_++1)}\right.\nonumber\\
&&	\left.+\frac{\omega_-}{\omega_++\omega_-}\frac{\Gamma(j+\omega_+)}{\Gamma(\omega_+)}\right)\nonumber\\
&&	\times\frac{\Gamma(\omega_++\omega_-+1)}{\Gamma(j+\omega_++\omega_-+1)}\frac{y^j}{j!},\\
&=&\sum_j\left(\frac{\omega_+}{\omega_++\omega_-}\frac{(j+\omega_+)\Gamma(j+\omega_+)}{\omega_+\Gamma(\omega_+)}\right.\nonumber\\
&&	\left.+\frac{\omega_-}{\omega_++\omega_-}\frac{\Gamma(j+\omega_+)}{\Gamma(\omega_+)}\right)\nonumber\\
&&	\times\frac{(\omega_++\omega_-)\Gamma(\omega_++\omega_-)}{(j+\omega_++\omega_-)\Gamma(j+\omega_++\omega_-)}\frac{y^j}{j!},\qquad\\
&=&\sum_j\frac{\Gamma(j+\omega_+)}{\Gamma(\omega_+)}\frac{\Gamma(\omega_++\omega_-)}{\Gamma(j+\omega_++\omega_-)}\frac{y^j}{j!}\\
&=&\Phi[\omega_+,\omega_++\omega_-;y],
\eeqn
as in Eqn.\ \ref{sri}.

The marginal $p_n$ is given by
\beq
\ip{n}{G} = \frac{1}{n!}\partial_x^n[G(x)]_{x=0}
\eeq
(cf.\ Eqn.\ \ref{cauchy}).  Using Eqn.\ \ref{sri} and the derivative of the confluent hypergeometric function,
\beq
\partial_y^n\Phi[\alpha,\beta;y] = \frac{\Gamma(n+\alpha)}{\Gamma(\alpha)}\frac{\Gamma(\beta)}{\Gamma(n+\beta)}\Phi[\alpha+n,\beta+n;y],
\eeq
one obtains Eqn.\ \ref{raj}.

\section*{APPENDIX C}

In this appendix, we fill in the details of the derivations of the equations of motion for the latter three of the four spectral bases discussed in Sec.\ \ref{bases}.

\subsection*{The $\ket{j,k_j}$ basis}

Projecting the conjugate state $\bra{j,k_j}$ onto Eqn.\ \ref{eom_2} (in which dummy indices $j$ and $k$ are changed to $j'$ and $k'$ respectively) gives
\beqn
0 &=& \sum_{j'k'}(j'+\rho k')\ip{j}{j'}\ip{k_j}{k'_{j'}}G_{j'k'}\nonumber\\
&&	+\sum_{j'k'}\bra{j}\bd_n\Gah\ket{j'}\ip{k_j}{k'_{j'}}G_{j'k'}\nonumber\\
\label{eom_2a}
&&	+\rho\sum_{j'k'}\bra{k_j}\bd_m\ket{k'_{j'}}\bra{j}\Deh(j')\ket{j'}G_{j'k'}.
\eeqn
From the orthonormality of states, the first term of Eqn.\ \ref{eom_2a} simplifies to
\beq
\sum_{k'}(j+\rho k')\ip{k_j}{k'_j}G_{jk'}
= (j+\rho k)G_{jk}.
\eeq
Recalling Eqn.\ \ref{overlap}, the product $\ip{k_j}{k'_{j'}}$ simplifies to
\beq
\ip{k_j}{k'_{j'}} = \frac{(-Q_{jj'})^{k-k'}}{(k-k')!}\theta(k-k'+1),
\eeq
with $Q_{jj'}=\qb_j-\qb_{j'}$, whereupon Eqn.\ \ref{eom_2a}, separating the part of its second term which is diagonal in $k$ from that which is subdiagonal and applying Eqn.\ \ref{bpa} to its third term, becomes
\beqn
0 &=& (j+\rho k)G_{jk}\nonumber\\
&&	+\sum_{j'}\Gamma_{j-1,j'}G_{j'k}\nonumber\\
&&	+\sum_{k'<k}\sum_{j'}\Gamma_{j-1,j'}\frac{(-Q_{jj'})^{k-k'}}{(k-k')!}G_{j'k'}\nonumber\\
\label{eom_2b}
&&	+\rho\sum_{k'<k}\sum_{j'}\Delta_{jj'}\frac{(-Q_{jj'})^{k-k'-1}}{(k-k'-1)!}G_{j'k'},
\eeqn
with $\Gamma_{jj'}$ as in Eqn.\ \ref{Gammajj} and
\beqn
\label{Deltaj1}
\Delta_{jj'}&=&\bra{j}\Deh(j')\ket{j'}=\bra{j}(\qb_{j'}-\qh)\ket{j'}\\
\label{Deltaj2}
&=&\bra{j}(\qb_j-\qh)\ket{j'}\\
&=&\sum_{n}\ip{j}{n}(\qb_j-q_n)\ip{n}{j'}
\eeqn
(where the orthonormality of $\ket{j}$ states is used in going from Eqn.\ \ref{Deltaj1} to Eqn.\ \ref{Deltaj2}).  Defining $\ell=k-k'$ and
\beq
V^\ell_{jj'} = \frac{(-Q_{jj'})^\ell}{\ell!},
\eeq
Eqn.\ \ref{eom_2b} can be written more compactly as Eqn.\ \ref{jkj_final}.

\subsection*{The $\ket{n,k_n}$ basis}

Projecting the conjugate state $\bra{n,k_n}$ onto Eqn.\ \ref{eom_3} (in which dummy indices $n$ and $k$ are changed to $n'$ and $k'$ respectively) gives
\beqn
0 &=& \sum_{n'k'}(g_{n'}+n'+\rho k')\ip{n}{n'}\ip{k_n}{k'_{n'}}G_{n'k'}\nonumber\\
&&	-\sum_{n'k'}g_{n'}\ip{n}{n'+1}\ip{k_n}{k'_{n'}}G_{n'k'}\nonumber\\
&&	-\sum_{n'k'}n'\ip{n}{n'-1}\ip{k_n}{k'_{n'}}G_{n'k'}\nonumber\\
\label{eom_3a}
&&	+\rho\sum_{n'k'}\Delta_{n'}\ip{n}{n'}\ip{(k-1)_n}{k'_{n'}}G_{n'k'},\qquad\quad
\eeqn
where $\Delta_n = \qb_n-q_n$.  Noting that, as in Eqn.\ \ref{overlap},
\beq
\label{Qoverlap}
\ip{k_n}{k'_{n\pm1}} = \frac{(-Q^\pm_n)^{k-k'}}{(k-k')!}\theta(k-k'+1),
\eeq
where $Q^\pm_n = \qb_n-\qb_{n\pm1}$, Eqn.\ \ref{eom_3a} becomes
\beqn
0 &=& (g_n+n+\rho k)G_{nk}\nonumber\\
&&	-g_{n-1}\sum_{k' \le k}\frac{(-Q^-_n)^{k-k'}}{(k-k')!}G_{n-1,k'}\nonumber\\
&&	-(n+1)\sum_{k'\le k}\frac{(-Q^+_n)^{k-k'}}{(k-k')!}G_{n+1,k'}\nonumber\\
\label{eom_3b}
&&	+\rho\Delta_nG_{n,k-1}.
\eeqn
Separating the parts of the second and third term that are diagonal in $k$ and defining $\ell=k-k'$ and
\beq
V^\pm_{n\ell} = \frac{(-Q^\pm_n)^\ell}{\ell!},
\eeq
Eqn.\ \ref{eom_3b} becomes Eqn.\ \ref{nkn_final}.

\subsection*{The $\ket{j,k_n}$ basis}

Substituting Eqn.\ \ref{pnm4} into Eqn.\ \ref{mastern} at steady state gives
\beqn
0&=&\sum_{jk}G_{jk} \left\{%\right.\nonumber\\
	g_{n-1}\ip{n-1,m}{j,k_{n-1}}\right.\nonumber\\
&&	+(n+1)\ip{n+1,m}{j,k_{n+1}}%\nonumber\\
	-(g_n+n)\ip{n,m}{j,k_n}\nonumber\\
&&	+\rho\left[q_n\ip{n,m-1}{j,k_n}%\right.\nonumber\\
	+(m+1)\ip{n,m+1}{j,k_n}\right.\nonumber\\
&&	\left.\left.-(q_n+m)\ip{n,m}{j,k_n}\right]\right\},
\eeqn
or, in terms of raising and lowering operators (cf.\ Eqns.\ \ref{apa}-\ref{ama}),
\beqn
0&=&\sum_{jk}G_{jk} \bra{n,m}\left\{%\right.\nonumber\\
	\ad_n\gh\ket{j,k_{n-1}}\right.\nonumber\\
&&	+\al_n\ket{j,k_{n+1}}%\nonumber\\
	-(\gh+\ad_n\al_n)\ket{j,k_n}\nonumber\\
&&	+\rho\left[\ad_m\qh\ket{j,k_n}%\right.\nonumber\\
	+\al_m\ket{j,k_n}\right.\nonumber\\
\label{jkn_tempa}
&&	\left.\left.-(\qh+\ad_m\al_m)\ket{j,k_n}\right]\right\}.
\eeqn
Using the definitions in Eqns.\ \ref{H}-\ref{b4}, Eqn.\ \ref{jkn_tempa} can be written as Eqn.\ \ref{jkn_temp2}.

Using Eqns.\ \ref{H0_4}-\ref{jknHeig}, Eqn.\ \ref{jkn_temp2} can be written
\beqn
0&=&-\bra{n,m}\sum_{j'k'}(j'+\rho k')\ket{j',k'_n}G_{j'k'}\nonumber\\
&&	-\bra{n,m}\sum_{j'k'}\bd_n\Gah\ket{j',k'_n}G_{j'k'}\nonumber\\
&&	-\bra{m}\rho\sum_{j'k'}(\qb_n-q_n)\ip{n}{j'}\bd_m\ket{k'_n}G_{j'k'}\nonumber\\
&&	+\bra{m}\sum_{j'k'}g_{n-1}\ip{n-1}{j'}\ket{\delta_-k'_n}G_{j'k'}\nonumber\\
\label{jkn_temp3a}
&&	+\bra{m}\sum_{j'k'}(n+1)\ip{n+1}{j'}\ket{\delta_+k'_n}G_{j'k'},
\eeqn
where
\beq
\ket{\delta_\pm k_n} = \ket{k_{n\pm1}}-\ket{k_n},
\eeq
and the dummy indices $j$ and $k$ have been changed to $j'$ and $k'$ respectively.  Using Eqn.\ \ref{Qoverlap} to note that
\beq
\ip{k_n}{\delta_\pm k'_n} = \frac{(-Q^\pm_n)^{k-k'}}{(k-k')!}\theta(k-k'),
\eeq
where $Q^\pm_n = \qb_n-\qb_{n\pm1}$, we multiply Eqn.\ \ref{jkn_temp3a} by $\ip{k_n}{m}$ and sum over $m$ to obtain
\beqn
0&=&-\bra{n}\sum_{j'}(j'+\rho k)\ket{j'}G_{j'k}\nonumber\\
&&	-\bra{n}\sum_{j'}\bd_n\Gah\ket{j'}G_{j'k}\nonumber\\
&&	-\rho\sum_{j'}(\qb_n-q_n)\ip{n}{j'}G_{j',k-1}\nonumber\\
&&	+\sum_{j'}g_{n-1}\ip{n-1}{j'}\sum_{\ell=1}^{k}V^-_{n\ell}G_{j',k-\ell}\nonumber\\
\label{jkn_temp4a}
&&	+\sum_{j'}(n+1)\ip{n+1}{j'}\sum_{\ell=1}^{k}V^+_{n\ell}G_{j',k-\ell},
\eeqn
in which we exploit the completeness of $\ket{m}$ states, i.e.\ $\sum_m\ket{m}\bra{m}=1$, and $V^\pm_{n\ell}$ is as in Eqn.\ \ref{Vnl}.  Multiplying Eqn.\ \ref{jkn_temp4a} by $\ip{j}{n}$, summing over $n$, and exploiting $\sum_n\ket{n}\bra{n}=1$ for the first two terms, we obtain Eqn.\ \ref{jkn_final}.

\section*{APPENDIX D}
In this appendix we explicitly solve for $G_{nk}$ in Eqn.\ \ref{trisimp} using the tridiagonal matrix, or Thomas \cite{thomas1949elliptic}, algorithm.  We start by identifying the subdiagonal, diagonal, superdiagonal, and right hand side elements of Eqn.\ \ref{trisimp}, respectively, as
\beqn
A_n &=& g \qquad\qquad\qquad\qquad\qquad\,\,\,\, (n = 1 \dots N), \\
B_n &=& -(\rho k + g + n) \qquad\qquad\,\,\,\,\,\, (n = 0 \dots N), \\
C_n &=& n+1 \qquad\qquad\qquad\qquad\,\,\,\,\, (n = 0 \dots N-1),\qquad\quad \\
R_n &=& -g\phi^-_k\delta_{nn_1}-n_1\phi^+_k\delta_{nn_0} \quad\,\, (n = 0 \dots N),
\eeqn
where $N$ is the cutoff in protein count $n$ and $n_1=n_0+1$. Auxiliary variables are defined iteratively as
\beqn
C'_0 &=& \frac{C_0}{B_0}, \\
\label{cpit}
C'_n &=& \frac{C_n}{B_n - C'_{n-1}A_n} \qquad (n = 1 \dots N-1),\\
R'_0 &=& \frac{R_0}{B_0}, \\
\label{dpit}
R'_n &=& \frac{R_n - R'_{n-1}A_n}{B_n - C'_{n-1}A_n} \qquad (n = 1 \dots N),
\eeqn
and the solution is obtained by backwards iteration with
\beqn
G^k_N &=& R'_N, \\
\label{Cit}
G^k_{n-1} &=& R'_{n-1} - C'_{n-1}G^k_n \qquad (n = N \dots 1)\qquad
\eeqn
(where $k$ has been moved from subscript to superscript for ease of reading).

Computing the first few terms of Eqn.\ \ref{cpit} reveals the pattern
\beq
\label{cp}
C'_n = -(n+1)\frac{\eta_n^k}{\eta_{n+1}^k},
\eeq
where
\begin{equation}
\label{eq:eta}
\eta_n^k = \sum_{i=0}^n \frac{n!}{i!(n-i)!}g^{n-i} \prod_{\ell=0}^{i-1} (\rho k+\ell),
\end{equation}
with the convention that $\prod_a^b [\cdot] = 1$ if $a>b$.
Note that since $\prod_{\ell=0}^{i-1} (\rho k+\ell)=\Gamma(\rho k+i)/\Gamma(\rho k)$, we may also use the integral representation of the Gamma function to write
\beqn
\eta_n^k&=&
\frac{1}{\Gamma(\rho k)}\sum_{i=0}^n \frac{n!}{i!(n-i)!}g^{n-i} 
\int_0^t dt \e{-t} t^{(\rho k+i-1)}\qquad\\
&=&
\frac{1}{\Gamma(\rho k)}\int_0^t dt \e{-t} t^{(\rho k-1)}
\sum_{i=0}^n \frac{n!}{i!(n-i)!}g^{n-i} t^i\\
&=&
\frac{1}{\Gamma(\rho k)}\int_0^t dt \e{-t} t^{(\rho k-1)}
(g+t)^n.
\eeqn

Using Eqn.\ \ref{dpit} it is immediately clear that
\beq
R'_{n<n_0} = 0.
\eeq
The first nonzero term is
\beq
R'_{n_0}
=\frac{n_1\phi_k^+}{gn_0}\frac{\eta^k_{n_0}}{\eta^k_{n_0-1}}\frac{1}{\epsilon^k_{n_0}-1},
\eeq
where we have defined
\beq
\epsilon_n^k = \frac{\rho k+g+n}{gn}\frac{\eta^k_n}{\eta^k_{n-1}}.
\eeq
Further iteration of Eqn.\ \ref{dpit} makes clear that
\beqn
R'_{n>n_0} &=& \frac{n_1}{g}f_k\prod_{i=n_0}^n\left(\frac{1}{i}\frac{\eta^k_i}{\eta^k_{i-1}}
	\frac{1}{\epsilon^k_i-1}\right)\\
&=& \frac{n_1}{g}\frac{(n_0-1)!}{n!}\frac{\eta^k_n}{\eta^k_{n_0-1}}f_k
	\prod_{i=n_0}^n\frac{1}{\epsilon^k_i-1},\quad
\eeqn
where
\beq
f_k = \phi^+_k-\frac{gn_0}{n_1}\frac{\eta^k_{n_0-1}}{\eta^k_{n_0}}(\epsilon^k_{n_0}-1)\phi_k^-.
\eeq

Computing the first few terms of Eqn.\ \ref{Cit} reveals that
\beqn
G^k_{n>n_0} &=& (\epsilon^k_n-1)R'_{n>n_0}F^k_n\\
&=& \frac{n_1}{g}\frac{(n_0-1)!}{n!}\frac{\eta^k_n}{\eta^k_{n_0-1}}f_kF^k_n
	\prod_{i=n_0}^{n-1}\frac{1}{\epsilon^k_i-1},\qquad
\eeqn
where
\beq
F^k_n = \sum_{i=0}^{N-n}\prod_{\ell=n}^{n+i}\frac{1}{\epsilon^k_\ell-1}.
\eeq
At the threshold Eqn.\ \ref{Cit} gives
\beq
G^k_{n_0} = \frac{n_1}{gn_0}\frac{\eta^k_{n_0}}{\eta^k_{n_0-1}}\frac{1}{\epsilon^k_{n_0}-1}
	(\phi^+_k+f_kF^k_{n_1})
\eeq
and since $R'_{n<n_0}=0$, the solution is easily completed using Eqn.\ \ref{Cit}, giving
\beqn
G_{n<n_0} &=& G_{n_0}\prod_{i=1}^{n_0-n}(-C'_{n_0-i})\\
&=& \frac{n_1}{g}\frac{(n_0-1)!}{n!}\frac{\eta^k_n}{\eta^k_{n_0-1}}
	\frac{\phi^+_k+f_kF^k_{n_1}}{\epsilon^k_{n_0}-1}.
\eeqn
These results are summarized in Eqns.\ \ref{Gana}-\ref{etaint}.

\begin{acknowledgments}
A.M.\ was supported by National Science Foundation Grant DGE-0742450.
A.M.\ and C.W.\ were supported by National Science Foundation Grant ECS-0332479.
A.M.W.\ was supported by the Princeton Center for Theoretical Science Fellowship and by Columbia's Professional Schools Diversity Short-Term Visiting Fellowship.
C.W.\ was supported by National Institutes of Health Grants 5PN2EY016586-03 and 1U54CA121852-01A1.
\end{acknowledgments}

\bibliographystyle{apsrev}
\bibliography{cspan3}

\end{document}